\documentclass[sn-standardnature]{sn-jnl}
\jyear{2024}%

\theoremstyle{thmstyleone}
\theoremstyle{thmstyletwo}%

\theoremstyle{thmstylethree}%
\usepackage{url}
\usepackage{xcolor}
\usepackage{anyfontsize}

\newcommand{\re}[1]{\textcolor{black}{#1}} 

\usepackage[mathlines]{lineno}
\usepackage{hyperref}
\begin{document}


\title[Main Text]{\vspace{-2cm} Mapping effective connectivity by virtually perturbing a surrogate brain}

\author[1]{\fnm{Zixiang} \sur{Luo}}
\author[1]{\fnm{Kaining} \sur{Peng}}
\author[1]{\fnm{Zhichao} \sur{Liang}}
\author[1]{\fnm{Shengyuan} \sur{Cai}}
\author[2]{\fnm{Chenyu} \sur{Xu}}
\author[1]{\fnm{Dan} \sur{Li}}
\author[3]{\fnm{Yu} \sur{Hu}}
\author[4]{\fnm{Changsong} \sur{Zhou}}
\author*[1]{\fnm{Quanying} \sur{Liu}}\email{liuqy@sustech.edu.cn}

\affil[1]{\orgdiv{Department of Biomedical Engineering}, \orgname{Southern University of Science and Technology}, \orgaddress{\city{Shenzhen}, \postcode{518055}, \country{China}}}

\affil[2]{\orgdiv{Department of Electrical and Computer Engineering}, \orgname{Iowa State University}, \orgaddress{\city{Ames}, \postcode{50011}, \state{Iowa}, \country{USA}}}

\affil[3]{\orgdiv{Department of Mathematics and Division of Life Science}, \orgname{The Hong Kong University of Science and Technology}, \orgaddress{\street{Hong Kong SAR}, \country{China}}}

\affil[4]{\orgdiv{Department of Physics, Centre for Nonlinear Studies}, \orgname{Hong Kong Baptist University}, \orgaddress{\street{Hong Kong SAR}, \country{China}}}

\abstract{
Effective connectivity (EC), indicative of the causal interactions between brain regions, is fundamental to understanding information processing in the brain. Traditional approaches, which infer EC from neural responses to stimulations, are not suited for mapping whole-brain EC in humans due to being invasive and having limited spatial coverage of stimulations. To address this gap, we present Neural Perturbational Inference (NPI), a data-driven framework designed to map EC across the entire brain. NPI employs an artificial neural network trained to learn large-scale neural dynamics as a computational surrogate of the brain. NPI maps EC by perturbing each region of the surrogate brain and observing the resulting responses in all other regions. NPI captures the directionality, strength, and excitatory/inhibitory properties of brain-wide EC. Our validation of NPI, using models having ground-truth EC, shows its superiority over Granger causality and dynamic causal modeling. Applying NPI to resting-state fMRI data from diverse datasets reveals consistent and structurally supported EC. \re{Further validation using a cortico-cortical evoked potentials dataset reveals a significant correlation between NPI-inferred EC and real stimulation propagation pathways.} By transitioning from correlational to causal understandings of brain functionality, NPI marks a stride in decoding the brain's functional architecture and facilitating both neuroscience research and clinical applications.
}

\maketitle

\newpage 

\section{Introduction}\label{sec1}

The brain operates as an intricate network of interconnected regions, which collaboratively processes external stimuli to generate behavior~\cite{Park2013Structural,deco2021revisiting}. Understanding the information flow between these regions \re{is} key to deciphering brain function~\cite{Park2013Structural,seguin2023brain}. While structural connectivity (SC) maps the brain's physical wiring and functional connectivity (FC) identifies statistical dependencies among neural activities, these measures fall short of illustrating the directional flow of information~\cite{Yeh2021Mapping, vandenHeuvel2010Exploring}. Effective connectivity (EC), \re{delineating} the causal interactions between brain regions, is thus essential for understanding information flow and critical in selecting target nodes for neuromodulation in brain disorder treatments~\cite{Schippers2010Mapping,manjunatha2024controlling}. 

EC is traditionally derived through neurostimulation experiments, such as optogenetics \cite{Kim2023Wholebrain, Randi2023Neural} or deep brain stimulation (DBS) \cite{Hollunder2024Mapping}. These methods involve perturbing specific brain regions and monitoring the resultant neural responses in other areas, thereby providing direct evidence of causality. However, such `perturbing and recording' procedures are invasive and do not scale well for whole-brain analysis. Computational approaches offer non-invasive alternatives but often suffer from inaccuracies, especially when applied at a whole-brain scale. Model-based methods, like Dynamic causal modeling (DCM), \re{heavily rely} on underlying model assumptions and are prone to biases from model mismatches~\cite{Friston2014DCM}. On the other hand, model-free methods such as Granger causality (GC) are adept at discerning the directionality of EC but struggle to accurately measure its strength or differentiate between excitatory and inhibitory influences~\cite{Li2018Causal}. Moreover, the interpretation of EC varies across computational frameworks, leading to ambiguity in the interpretation of EC inferred from computational and experimental approaches.

The advent of big data in neuroscience, propelled by advanced imaging and electrophysiological techniques, has facilitated the use of artificial neural networks (ANN) to analyze complex neural data~\cite{Liang2022Online,abrol2021deep}. \re{Recurrent neural network models have been employed to learn temporal dynamics of brain signals and infer EC directly from the learned weight matrices~\cite{Perich2020Inferring, tu2019state}. While these models can capture brain dynamics, there is no guarantee that the learned weights reflect the underlying EC, particularly when the model's assumptions do not align with the brain's underlying dynamics and when dealing with a large number of regions~\cite{Das2020Systematic}.} Perturbation analysis in ANN presents a promising avenue for investigating causality, where modulating input variables and observing subsequent output changes allow for the elucidation of causal relationships tied to specific inputs and their effects~\cite{ivanovs2021perturbation, dong2023causal}. \re{Such perturbational approaches are conceptually similar to using stimulation-evoked potentials to infer EC in neuroscience, aiming to delineate causal connections~\cite{Veit2021Temporal, Ozdemir2020Individualized}}. Inspired by this parallel, our study integrates perturbation-based experiments into a data-driven framework, revealing the brain causality at a whole-brain level.

In this study, we present the Neural Perturbational Inference (NPI) technique for non-invasively mapping whole-brain EC. NPI utilizes an ANN that learns the brain dynamics as a surrogate brain. After the ANN is well-trained to capture the brain-wide neural dynamics, systematically perturbing the trained ANN yields a map of causal relationships among all brain regions. It delineates the directionality, strength, and excitatory/inhibitory properties of whole-brain causal interactions. The effectiveness of NPI is validated on a variety of generative models with established ground-truth EC. \re{NPI shows a remarkable match with cortico-cortical evoked potentials, validating its accuracy in reflecting real causal interactions in the brain.} NPI holds promise for advancing the understanding of brain information flow and the clinical treatment of neurological disorders.

\begin{figure}[t!]
\centering
\includegraphics[width=1\linewidth]{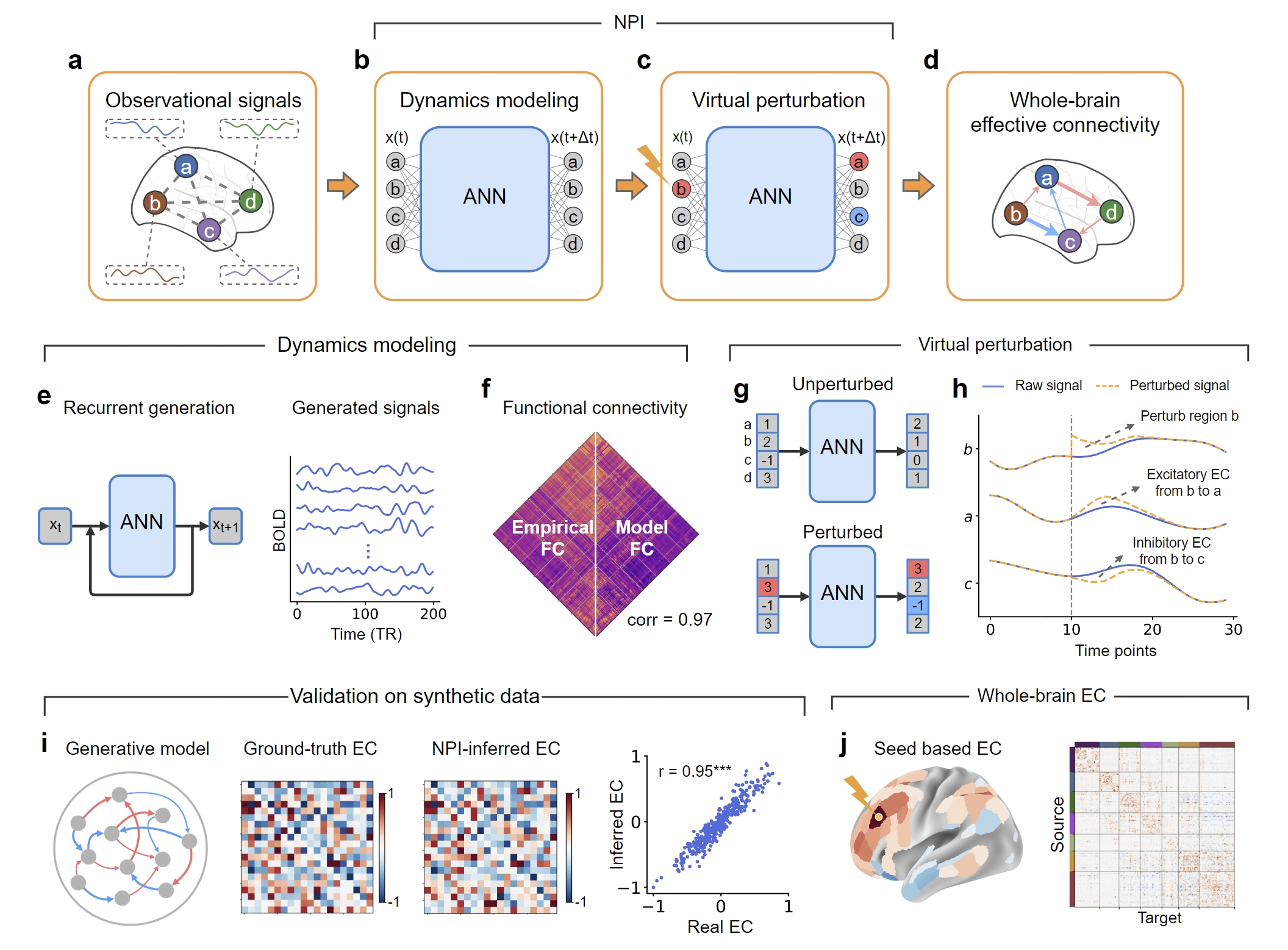}
\caption{\textbf{Neural Perturbational Inference (NPI) maps effective connectivity (EC) by virtually perturbing a surrogate brain.} 
\textbf{a}, Schematic of the brain network and the recorded neural signals of each brain region, from which the EC values among regions are inferred.
\textbf{b}, A surrogate brain, an artificial neural network (ANN), is trained to learn brain dynamics. It can then replace the real brain to be perturbed. ANN is optimized for predicting the subsequent brain state given the \re{previous brain states}.
\textbf{c}, After training, ANN is systematically perturbed to infer EC. After perturbing one region, the magnitudes of the perturbation-induced responses refer to a one-to-all EC.
\textbf{d}, The all-to-all EC can be inferred by perturbing the ANN region by region. This EC is a brain-wide map of causal influences that shows directionality, strength, and excitatory/inhibitory distinction.
\textbf{e}, Recurrently feeding the result of prediction back as input to ANN produces the generated neural signals. 
\textbf{f}, The model FC and empirical FC are respectively calculated from generated \re{individual} BOLD signals and empirical \re{individual} BOLD signals, respectively, and then averaged across 800 subjects. The model FC and the empirical FC are highly correlated ($r=0.97$, $p < 10^{-3}$), indicating that the trained ANN as a surrogate brain captures the inter-regional \re{relationships} of the real brain.
\textbf{g}, Perturbation is applied as an increase of neural signal at a selected region. Changes in the predicted responses of target regions, induced by perturbed input versus baseline input, reflect the EC from the source to the target regions. The effect of perturbation is indicated by a change in color: red represents an increase in neural signal relative to the unperturbed state, while blue denotes a decrease. 
\textbf{h}, Perturbing region $b$ caused an increase of subsequent activity in region $a$ and a decrease of subsequent activity in region $c$, indicating an excitatory EC from $b$ to $a$ and an inhibitory EC from $b$ to $c$. 
\textbf{i}, Effectiveness of NPI is validated on generative models with known ground-truth EC. The NPI-inferred EC recovers the strength, directionality, and excitatory/inhibitory properties of EC with a high accuracy.
\textbf{j}, NPI applied to resting-state fMRI data gives the whole-brain EC \re{from source to target regions}.
}
\label{fig:framework}
\end{figure}

\section{Results}\label{sec2} 
\subsection{Neural Perturbational Inference}

NPI is a framework that non-invasively infers EC from neural signals (Fig.~\ref{fig:framework}a-d). Conceptually, NPI is similar to perturbing the real brain through neurostimulation, but it uses an ANN as a surrogate brain to replace the real brain, which enables efficient whole-brain perturbation and observation.

From brain imaging or electrophysiological recordings, the collective neural activities of multiple brain regions are easily available, but how these regions interact to process information is unclear (Fig.~\ref{fig:framework}a). NPI aims to infer EC among regions for the entire brain, which are directed causal connections. This study implemented the ANN as a multi-layer perceptron (MLP; \re{Supplementary Fig. 1}). The ANN in NPI can be implemented as different predictive models as long as the model can learn brain dynamics and capture inter-region relationships \re{(Supplementary Fig. 2, Supplementary Note 1,2)}. In addition to the MLP network, we tested various surrogate models (e.g., CNN, RNN, VAR) to assess their performance in signal prediction, FC reproduction, and EC inference (\re{Supplementary Table 1}). The results show that the NPI framework remains robust across different ANN architectures. The ANN is trained to predict the brain state at the next time step based on the brain states of \re{the preceding three time steps} by minimizing the one-step-ahead prediction error (Fig.~\ref{fig:framework}b). To validate the ability of ANN to capture the interaction relationships between brain regions. We recursively fed the predicted output into the ANN and generated the synthetic signals (Fig.~\ref{fig:framework}e). \re{On human BOLD data, the FC calculated from the synthetic BOLD signals (model FC) and the empirical BOLD signals (empirical FC) are compared, both of which are averaged across 800 subjects in the HCP dataset.} The model FC and empirical FC are strongly correlated ($r=0.98$, $p < 10^{-3}$), suggesting ANN captures \re{complex inter-region relationship in the brain, which is crucial for the EC inference} (Fig.~\ref{fig:framework}f). This suggests that the trained ANN can serve as a surrogate brain for virtual perturbations.

The trained ANN is fixed and treated as a surrogate model for the brain. We then applied virtual perturbations to each node of the ANN, with each node representing a brain region (Fig.~\ref{fig:framework}c). The perturbation is implemented as an impulse increase to the signal at the selected node \re{at time $t$} (Fig.~\ref{fig:framework}g). The ANN takes both perturbed input and baseline input to predict \re{subsequent neural activities $x(t+1)$}. Changes in the predicted responses of target regions — when comparing perturbed input to baseline input — reflect the EC from the source region (the perturbed region) to the target regions. Increased or decreased activity in the target regions indicates excitatory or inhibitory EC, respectively (Fig.~\ref{fig:framework}h). Systematically perturbing each node in the ANN reveals the all-to-all EC (Fig.~\ref{fig:framework}d), characterizing the directionality, strength, and excitatory/inhibitory properties of causal influences among brain regions. We show that this systematic perturbation is interpreted as the Jacobian matrix of the trained ANN (Supplementary Note 4, Extended Data Fig.~\ref{fig:jacobian}), which quantifies how a small input to one node can positively or negatively influence the next states of other nodes.

To validate the effectiveness of NPI, we applied it to data generated by pre-defined generative models with established ground-truth EC (Fig.~\ref{fig:framework}i). We quantify the EC inference performance by comparing the NPI-inferred EC with the ground-truth EC. When applied to real rs-fMRI datasets, NPI can reveal seed-based EC and the whole-brain EC, uncovering the distribution of EC both within and across functional brain networks (Fig.~\ref{fig:framework}j).

\subsection{Validation of NPI on generative models}

\begin{figure}[!htbp]
\centering
\includegraphics[width=1\linewidth]{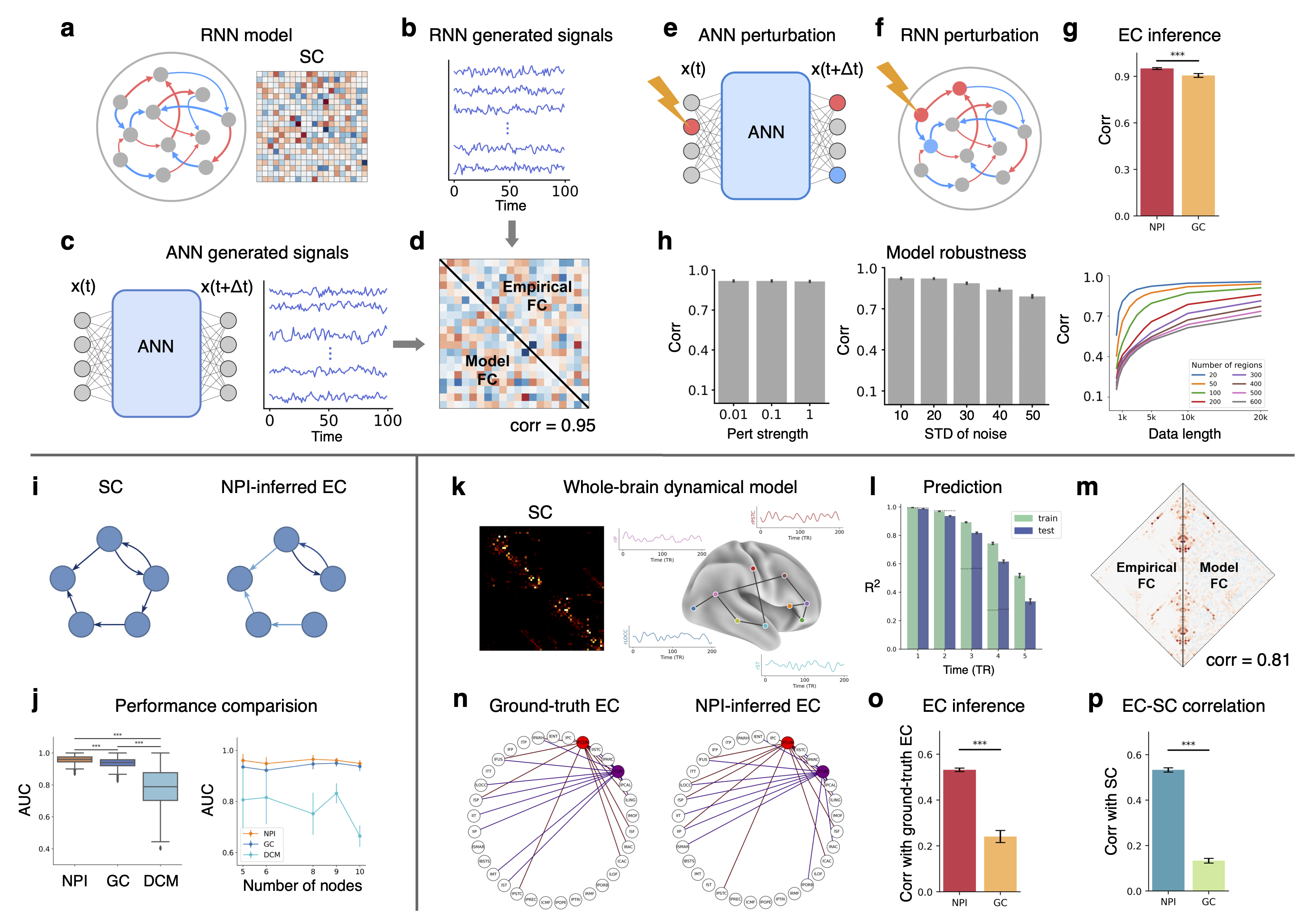}
\caption{\textbf{Validation of NPI on generative models.}
\textbf{a}, to \textbf{h}, Validation of NPI on synthetic data generated by ground-truth recurrent neural networks (RNN).
\textbf{a}, Structural connectivity (weight matrix) of the RNN model.
\textbf{b}, RNN-generated synthetic neural signals used to train ANN and compute empirical FC.
\textbf{c}, Left: An ANN is trained to predict subsequent states of the RNN-generated data. Right: Recursively running the ANN can generate signals that capture RNN dynamics. ANN-generated signals are used to compute model FC.
\textbf{d}, Model FC is strongly correlated with empirical FC ($r=0.95, p < 10^{-3}$, $t$-test for correlation).
\textbf{e}, NPI maps EC by virtual `perturb and record' protocol.
\textbf{f}, Perturbing RNN model produces the ground-truth EC.
\textbf{g}, NPI-inferred EC is strongly correlated with ground-truth EC ($r=0.95, p < 10^{-3}$, $t$-test for correlation), better than GC-inferred EC ($p < 10^{-3}$, Wilcoxon signed-rank test). Error bars represent standard deviation.
\textbf{h}, The robustness of NPI's inference performance relates to perturbation magnitudes, noise intensities, and data lengths. Pearson's correlation between NPI-inferred EC and ground-truth EC is calculated. Error bars and shadows represent standard deviation.
\textbf{i} to \textbf{j}, Validation of NPI on an open source synthetic fMRI dataset.
\textbf{i}, NPI-inferred EC captures the characteristics of the SC from a public dataset proposed by Sanchez-Romero et al.~\cite{Sanchez-Romero2019Estimating}. 
\textbf{j}, NPI outperforms GC and DCM in classifying the presence or absence of possible SC connections ($p < 10^{-3}$, Wilcoxon signed-rank test), and the advantage remains consistent regardless of the number of nodes in the SC. Error bars represent standard deviation.
\textbf{k} to \textbf{p}, Validation of NPI on WBM-generated synthetic data.
\textbf{k}, fMRI signals are simulated by a large-scale dynamic model based on an SC matrix derived from DSI data of the human brain.
\textbf{l}, The prediction performance of the ANN remains significant over multiple time steps. Error bars represent standard deviation. \re{The dotted line represents the performance of the univariate auto-regression baseline model.}
\textbf{m}, The surrogate brain successfully reproduces the empirical FC ($r=0.81$, $p < 10^{-3}$, $t$-test for correlation).
\textbf{n}, NPI-inferred EC closely resembles ground-truth EC, derived by perturbing WBM directly. \re{The strongest 40\% output EC from two median-performing nodes in the left hemisphere to the rest of the nodes in the left hemisphere are illustrated.}
\textbf{o}, \textbf{p}, NPI outperforms GC in capturing the characteristics of both \re{whole-brain} EC (o) and SC (p) ($p < 10^{-3}$, Wilcoxon signed-rank test). Error bars represent standard deviation.
}
\label{fig:validation}
\end{figure}


We first validated the capability of NPI by applying it to infer EC from synthetic data generated by models with established ground-truth EC (see Methods). We used three simulated datasets: synthetic data generated by ground-truth recurrent neural network (RNN) models, a public synthetic BOLD dataset with few brain regions, and synthetic BOLD data using a whole brain model (WBM). To derive the ground-truth EC, we used the 'perturb and record' protocol directly on the generative models. We assessed NPI's inference performance by comparing this ground-truth EC with the EC inferred by NPI. 

NPI was firstly applied to infer EC from a RNN with a pre-defined weight matrix serving as SC, where the entries were drawn from a Gaussian distribution centered at zero (Fig.~\ref{fig:validation}a). The neural signals are then synthesized by executing the RNN (Fig.~\ref{fig:validation}b).
An ANN is fitted to the signals generated by the RNN as a surrogate. The ANN's ability to learn the non-linear RNN system dynamics is evidenced by its successful generation of synthetic signals when its output is recursively fed back into the system (Fig.~\ref{fig:validation}c). FC derived from the ANN-synthesized signals demonstrated a strong correlation with FC directly calculated from the RNN-generated signals, implying ANN's proficiency in capturing the RNN's inter-regional dynamics (Fig.~\ref{fig:validation}d). 

To derive EC, perturbations are then applied to the trained ANN (Fig.~\ref{fig:validation}e, Supplementary Fig. 2,3). The RNN's intrinsic EC, obtained through perturbing the ground-truth RNN directly, is used as ground-truth EC (Fig.~\ref{fig:validation}f). We calculated the correlation between NPI-inferred EC and ground-truth EC, as well as the correlation between GC-inferred EC and ground-truth EC (Supplementary Note 5). The results show a strong alignment between NPI-inferred EC and the ground-truth EC, with correlation coefficient of $r=0.95$, outperforming GC (Fig.~\ref{fig:validation}g, Supplementary Fig. 4). The NPI-inferred EC also demonstrates a strong correlation with the SC of RNN, which serves as the anatomical foundation for EC (Supplementary Fig. 4). EC does not perfectly align with SC due to the inherent nonlinearity of brain dynamics and signal noise, the correlation between EC and SC is significantly stronger than that between FC and SC (Supplementary Table 2). This is likely because FC lacks directionality and suffers from spurious connectivity~\cite{power2012spurious}.
To evaluate the robustness of the NPI, we conducted comprehensive analyses, including applying a range of perturbation intensities to the ANN, varying levels of systemic noise to the RNN model, and varying data lengths and RNN sizes (Fig.~\ref{fig:validation}h). The results showed that NPI's EC inference performance remains stable across different perturbation magnitudes and experiences only a slight decline with increasing noise levels, demonstrating the method's robustness. In scenarios with varying data lengths and RNN sizes, we found that larger datasets are crucial for reliable EC inference on larger networks.


To examine the NPI's efficacy on BOLD signals and on networks with different structures, we applied NPI to a public synthetic dataset containing BOLD dynamics generated from nine different underlying SC structures~\cite{Sanchez-Romero2019Estimating} (Fig.~\ref{fig:validation}i, Extended Data Fig.~\ref{fig:opendata}a). This dataset, widely used in validating EC inference algorithms, features binary SC and simulates neural firing rates subsequently converted into BOLD signals through a hemodynamic response function (see Methods). 
For this dataset, as the ground-truth EC is unavailable, we evaluated the performance of EC inference using the Area Under the Receiver Operating Characteristic Curve (AUC) by classifying the presence or absence of each possible SC connection after binarizing the NPI-inferred EC. We show that NPI achieved an AUC close to 1, surpassing GC and DCM (Fig.~\ref{fig:validation}j). Across all nine SC configurations, NPI significantly outperformed both GC and DCM (Extended Data Fig.~\ref{fig:opendata}), demonstrating its precision and reliability in mapping EC across diverse connection topographies and model structures. 

Inferring EC from a large-scale network poses challenges for conventional methods like DCM. To validate NPI's effectiveness in large-scale EC inference, we applied NPI to the synthetic BOLD data generated from a whole-brain model (WBM) with 66 nodes (see Methods). Specifically, we utilized neuroanatomical connectivity data obtained via Diffusion Spectrum Imaging (DSI) as the underlying SC matrix. The BOLD time series were then generated by a neurodynamic model (Fig.~\ref{fig:validation}k). Despite a decline in multi-step prediction accuracy (Fig.~\ref{fig:validation}l, Supplementary Fig. 7), the FC of the ANN-generated signals shows a strong correlation with the FC of the WBM-simulated signals (Fig.~\ref{fig:validation}m), highlighting the ANN's effectiveness in capturing the inter-regional relationships.
Ground-truth EC from the WBM was obtained by perturbing each node and \re{observing} the resulting responses. The NPI-inferred EC not only shows a strong correlation with the ground-truth EC but also closely aligns with the underlying SC (Fig.~\ref{fig:validation}n, Supplementary Fig. 5,6). Furthermore, NPI-inferred EC more accurately reflects both the ground-truth EC and SC compared to EC inferred by GC (Fig.~\ref{fig:validation}o,p, $p<10^{-3}$, Wilcoxon signed-rank test), establishing NPI as a robust and reliable method for EC estimation in complex brain networks. On this dataset, we tested the performance of different surrogate models and found MLP gives the best FC reproduction and EC inference performance (Supplementary Tables 1, and 2). We thus use the MLP to be the surrogate model for inferring EC from real data.

\subsection{Human EBC inferred by NPI}

\begin{figure}[htbp]
\centering
\includegraphics[width=1\linewidth]{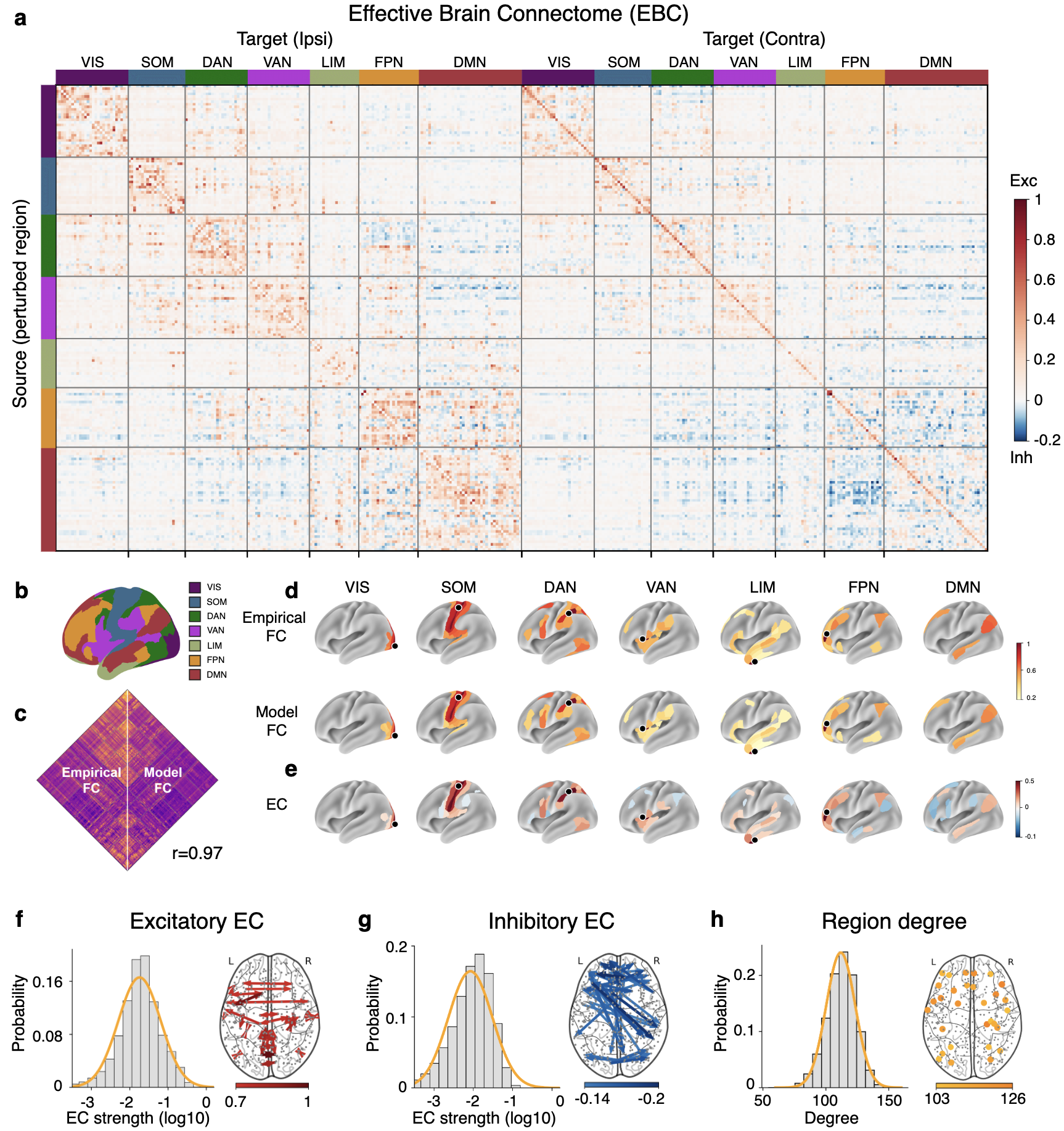}
\caption{\textbf{Human EBC inferred by NPI.}
\textbf{a}, The averaged EBC of 800 subjects, with regions organized according to functional networks. Each row represents the EC from a source region in the left hemisphere to the entire cortex with the maximum response scaled to $1.0$. The human EBC for the entire brain is shown in Extended Data Fig.~\ref{fig:wholeec}. 
\textbf{b}, \re{Cortical areas are assigned to seven functional resting-state networks.}
\textbf{c}, \re{FC derived from both the model-generated and empirical BOLD signals demonstrates a strong positive correlation ($r=0.97$).}
\textbf{d}, Maps of seed-based FC analysis \re{on the empirical BOLD data (i.e., empirical FC) and the ANN-generated BOLD data (i.e., model FC)}. The seed is set in seven resting-state networks, respectively. The seed region is indicated with a black dot on each map, except for DMN where the seed is inside the brain. FC entries with the top 10\% strength in each resting-state network are plotted.
\textbf{e}, Maps of NPI-inferred EC with a seed region in seven resting-state networks respectively. Excitatory EC values are shown in red. Inhibitory EC values are shown in blue. EC entries with top 10\% strength are plotted.
\textbf{f}, Left: The strength of excitatory EC follows a log-normal distribution, as demonstrated by the fitting curve of log-transformed EC for a Gaussian distribution. Right: The 50 strongest excitatory EC.
\textbf{g}, The same as (f) for inhibitory EC.
\textbf{h}, Left: The degree distribution of regions from the EC binarized by a threshold of 80\% strength. The degree of a region is calculated as the average of the in-degree and out-degree of that region. Right: 30 brain regions with the largest degree after binarizing EBC.
\textit{Abbrev.}: VIS, visual network; SOM, somatomotor network ; DAN, dorsal attention network; VAN, ventral attention network; LIM, limbic network; FPN, frontoparietal network; DMN, default mode network.}
\label{fig:ebc}
\end{figure}

\re{We applied NPI to resting-state fMRI (rs-fMRI)} data from 800 subjects in the Human Connectome Project (HCP) dataset parcellated using the Multi-Modal Parcellation atlas with 360 regions (Supplementary Table 5)~\cite{VanEssen2013WUMinn,Glasser2016Multimodal}. The \re{individualized} ANN was trained on the rs-fMRI data of each subject (see Methods). \re{Using the signals from the previous three steps to predict the next step yielded slightly better performance compared to using only the signals from the previous step as input (Supplementary Fig. 8). Therefore, we used the 3-step input MLP model for the following analysis. The trained ANN can be treated as an individualized surrogate model.} The group-level FC calculated from the real BOLD signals (i.e., empirical FC) and the ANN-generated BOLD signals (i.e., model FC) have a strong positive correlation (r=0.97, Fig.~\ref{fig:ebc}c) and share similar spatial patterns (Fig.~\ref{fig:ebc}d), suggesting that \re{the trained ANN captures the complex inter-regional interactions of the biological brain}.

After the surrogate model was trained, we applied systematic perturbations to each individualized surrogate model to obtain the whole-brain EC, which we call the effective brain connectome (EBC). \re{We first obtained the individualized EBC by perturbing the individualized surrogate model and then calculated the group-level EBC (i.e., Human EBC) by averaging the EBC across 800 subjects (Fig.~\ref{fig:ebc}a).} The positive entries indicate excitatory EC and negative entries indicate inhibitory EC. The brain regions are assigned to seven functional networks (i.e., visual network (VIS), somatomotor network (SOM), dorsal attention network (DAN), ventral attention network (VAN), limbic network (LIM), frontoparietal network (FPN), and default mode network (DMN)) according to Yeo et al.~\cite{ThomasYeo2011organization} (Supplementary Table 3, Fig.~\ref{fig:ebc}b). Seed-based EC is then analyzed to examine the topographic organization of functional networks. The top 10\% excitatory and top 10\% inhibitory output EC from seeds in six functional brain networks are plotted, showing a similar structure as networks defined by FC and better reflects how seed regions inhibit other parts across \re{the} whole brain (Fig.~\ref{fig:ebc}e).

The majority of EC have small and near-zero strengths, with a few having very large strengths. The distribution shows a long-tail property. We fit the strengths to four hypothesized distributions: log-normal, normal, exponential, and inverse Gaussian. According to the Akaike information criterion (AIC), the log-normal distribution is the best fit for both excitatory and inhibitory EC (Fig.~\ref{fig:ebc}f,g, Supplementary Table 4). It is consistent with the distribution of SC found in experimental studies using tract-tracing techniques involving mice and macaques~\cite{Oh2014mesoscale, Markov2014Weighted}. The log-normal distributions of excitatory and inhibitory EC are reproducible under the Automated Anatomical Labeling (AAL) parcellation (Supplementary Fig. 9). The excitatory EC has stronger strength than inhibitory EC. When the maximum strength of excitatory EC is scaled to 1, inhibitory EC has a maximum strength of $0.22$. The strongest excitatory EC are mostly intra-network connections, either intra-hemisphere or inter-hemisphere (Fig.~\ref{fig:ebc}f, Supplementary Fig. 10). The strongest inhibitory EC are mostly inter-network connections and are all inter-hemisphere connections (Fig.~\ref{fig:ebc}g, Supplementary Fig. 10). The degree of a node refers to the number of connections it has with other nodes in the network and can be used to measure the centrality or importance of that node in the network. We binarize the EBC at a threshold of $80\%$ absolute EC strengths (0.06). The EC with absolute strengths below the threshold are set to 0, while the rest are set to 1. The excitatory and inhibitory EC are not differentiated in binarized EBC. Since EC is directed and thus asymmetric, the in-degree of a node is different from the out-degree. In binarized EBC, most of the EC are bidirectional (73\%), consistent with previous findings on SC~\cite{Felleman1991Distributed}. Regions with the largest averaged in-out degrees are dispersed across the cortex in several functional networks (Fig.~\ref{fig:ebc}f). \re{Moreover, we reported the human EBC with 100, 200, ..., up to 1000 regions parcellated from the Schaefer atlas~\cite{schaefer2018local}. Results showed that EC inferred with atlases with different numbers of regions are highly stable and reliable (Extended Data Fig.~\ref{fig:atlas1000}).}

\begin{figure}[tbhp]
\centering
\includegraphics[width=1\linewidth]{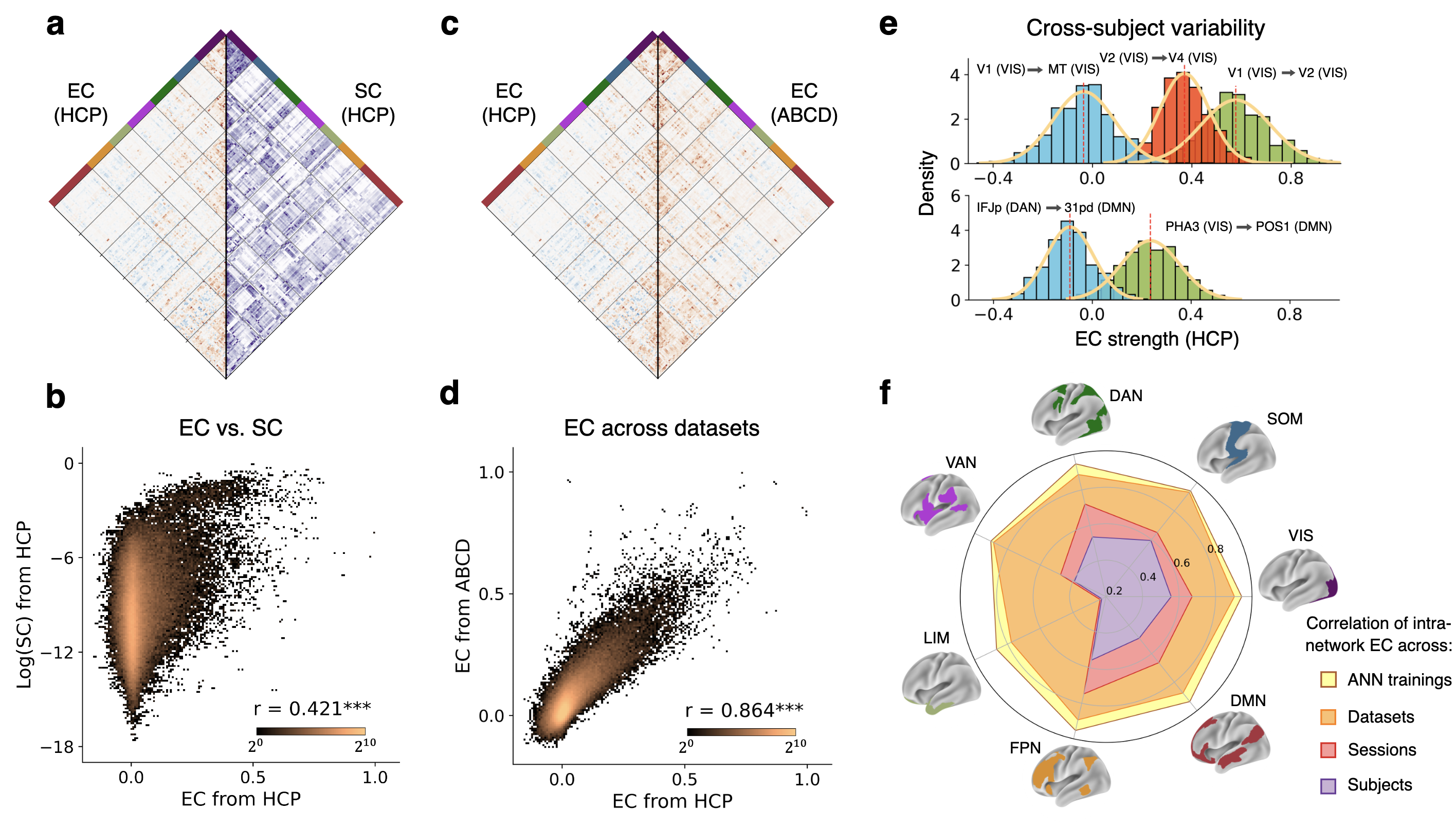}
\caption{\textbf{NPI-inferred EC is robust and aligns with structural connectivity.}
\textbf{a}, NPI-inferred EC (left) and SC derived by diffusion spectrum imaging (right) of the HCP dataset. 
\textbf{b}, Quantitative relationship between NPI-inferred EC and SC. The strengths of EC (absolute value) \re{show} a high correlation with strengths of SC, suggesting the association between structural substrates and effective neural communications ($r=0.42$, $p < 10^{-3}$, $t$-test for correlation).
\textbf{c}, EC inferred from the HCP dataset (left) and ABCD dataset (right).
\textbf{d}, Quantitative relationship between EC inferred from the HCP dataset and those from the ABCD dataset, demonstrating consistency in EC estimation across distinct datasets ($r=0.87$, $p < 10^{-3}$, $t$-test for correlation).
\textbf{e}, Cross-subject variability of EC strengths from 800 subjects.
Top: within-network EC variations, highlighting inter-subject variability for EC within the visual network (VIS). We show EC from V1 to V2 (green, mean=0.58, std=0.14), from V2 to V4 visual area (red, mean=0.37, std=0.10), and from V1 to MT (blue, mean=-0.04, std=0.12). 
\re{Bottom: cross-network EC variations. Here, we focus on the EC from VIS and the dorsal attention network (DAN) to the default mode network (DMN). Our results show a positive EC from PHA3 in VIS to POS1 in DAN (green, mean=0.24, std=0.12) and a negative EC from IFJp in DAN to 31pd in DMN (blue, mean=-0.09, std=0.10).}
\textbf{f}, Correlation of intra-network EC of seven functional networks across ANN trainings, sessions, subjects, and datasets. ANN trainings refers to running NPI twice using the data from the same individual with results averaged across 800 subjects. Sessions refers to splitting and training on each half of the individual data with results averaged across 800 subjects. Subjects refers to cross-subject EC correlation from 800 subjects in the HCP dataset. Datasets refers to the consistency between EC inferred from the HCP and the ABCD datasets. \re{\textit{Abbrev.}: V1, primary visual cortex; V2, secondary visual cortex; MT, middle temporal visual area; PHA3, Parahippocampal Area 3; POS1, Parieto-Occipital Sulcus area 1; IFJp, inferior frontal junction posterior part; 31pd, Posterior Dorsal area 31.}
}
\label{fig:robustness}
\end{figure}

\subsection{EBC is robust and congruent with structural basis}

To assess the reliability of EC inferred from fMRI data, we examined the relationship between EC and its structural foundation, derived from DSI. Our analysis revealed a strong correlation between EC and SC, confirming that the brain's anatomical structure plays a key role in shaping the pathways of functional neural communication (Fig.~\ref{fig:robustness}a,b). To further evaluate the robustness and consistency of EC inferred by NPI, we extended our analysis to the Adolescent Brain Cognitive Development (ABCD) dataset~\cite{saragosa2022practical}. The alignment of population-averaged EBC between the HCP and ABCD datasets highlights NPI's robust applicability across datasets and validates its potential for generalization (Fig.~\ref{fig:robustness}c,d).

We then tested the inter-subject variability of inferred EC. The inter-subject variability of within-network and \re{cross-network EC} are in the same range (Fig.~\ref{fig:robustness}e). Among all the EC pairs, 55\% of EC connections are significantly different from zero across 800 subjects, indicating a consistent deviation from a null hypothesis of no connection (Bonferroni corrected, Supplementary Fig. 11). To determine whether NPI-mapped EC depends on the variability of ANN training, we performed NPI twice on each subject from the HCP dataset, training two ANNs with different initializations. We assessed the consistency of EC obtained by perturbing two trained ANNs (termed as `ANN trainings' in Fig.~\ref{fig:robustness}f, yellow), showing that NPI-inferred EC is robust across ANN training. To distinguish intrinsic individual variability from potential noise introduced by the method, we conducted assessments of cross-session, inter-subject, and inter-dataset variability (termed as `Sessions', `Subjects' and `Datasets' in Fig.~\ref{fig:robustness}f, Supplementary Fig. 15). In the cross-session assessment, we split each individual’s data in half and examined the consistency of EC between the two halves. We found that cross-session EC exhibits a higher correlation than inter-subject EC, suggesting that NPI-inferred EC from the same subject is stable across sessions and NPI-inferred EC captures individual variability. \re{The limbic network exhibited the lowest reliability, likely due to the low signal-to-noise ratio of fMRI in this region~\cite{Liu2020Individual,ThomasYeo2011organization}.} Overall, our results suggest that NPI can reliably capture the general EBC patterns across datasets and effectively characterize the EC profiles of individual brains.

\subsection{NPI supports clinical applications}

\begin{figure}[H]
\centering
\includegraphics[width=0.9\linewidth]{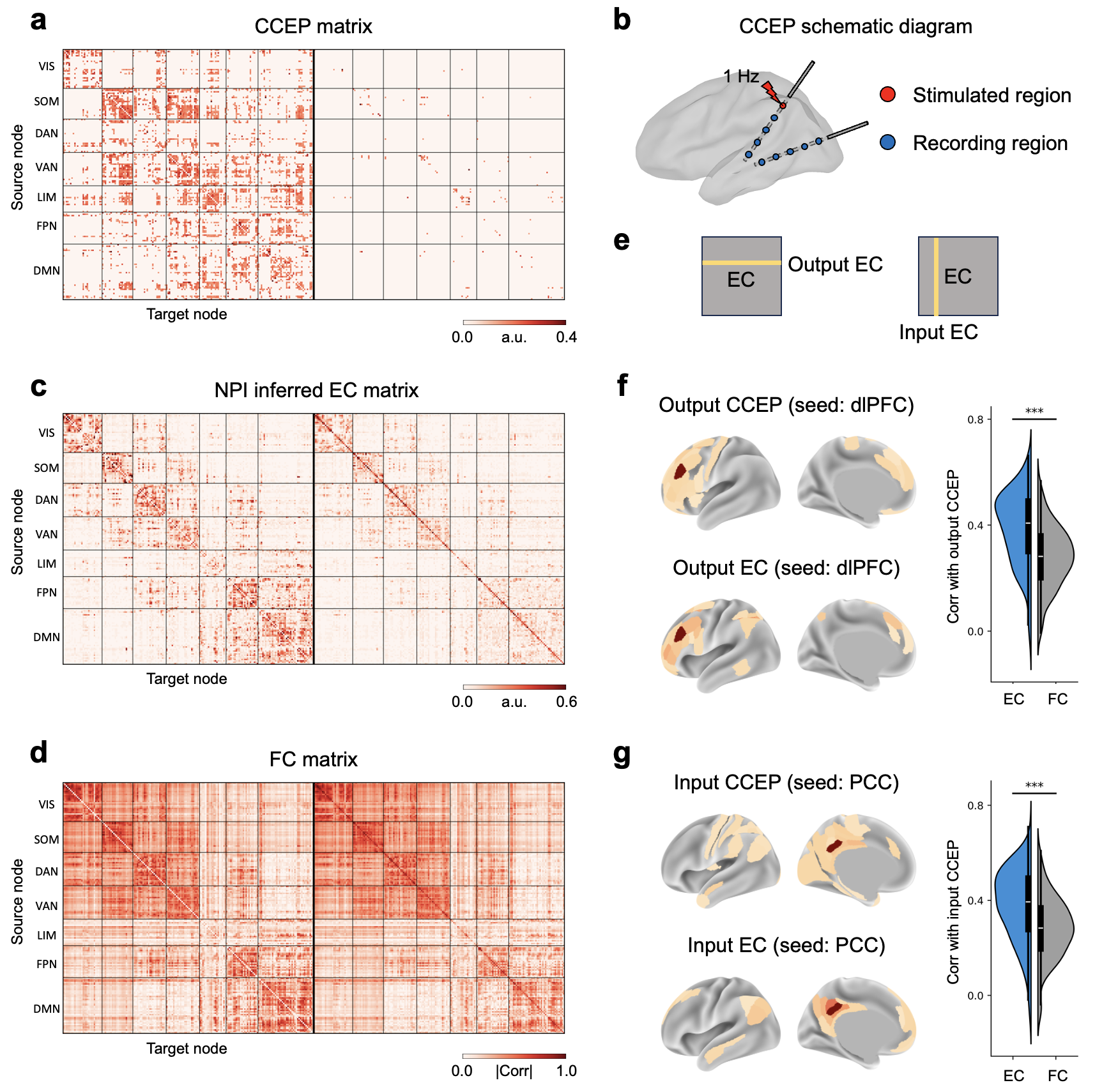}
\caption{\re{\textbf{Validation of NPI-inferred EC with cortico-cortical evoked potentials (CCEP).}
\textbf{a}, Group-level CCEP matrix from the Functional Brain Tractography (F-TRACT) project involving 613 patients, depicting evoked responses from the left hemisphere to the entire brain. 
\textbf{b}, Schematic representation of the CCEP experimental setup, showing invasive stimulation and recording locations.
\textbf{c,d}, Matrices of group-level NPI-inferred EC and empirical FC, respectively, from the HCP dataset, each involving 800 subjects and organized in congruence with the CCEP matrix. 
\textbf{e}, Left: A row of the CCEP matrix or NPI-inferred EC matrix represents the output CCEP or EC from a source region to all other regions. Right: A column of the CCEP matrix or NPI-inferred EC matrix represents the input CCEP or EC from all other regions to a target region.
\textbf{f}, Left: output CCEP and output EC from dlPFC show large similarity. Right: correlations between rows of EC and CCEP matrix and between rows of FC and CCEP matrix.
\textbf{g}, Left: input CCEP and input EC to PCC shows large similarity. Right: correlations between columns of EC and CCEP matrix and between columns of FC and CCEP matrix.
}}
\label{fig:ccep}
\end{figure}

\re{To validate the NPI’s potential for clinical applications, we examined the consistency between the spatial distribution of NPI-inferred EBC and neurostimulation-induced neural responses. We utilized an open-source cortico-Cortical Evoked Potentials (CCEP) dataset (Fig.~\ref{fig:ccep}a) from the Functional Tractography (F-TRACT) project~\cite{lemarechal2022brain}, which includes intracortical stimulation and intracerebral stereoencephalographic (SEEG) recordings in epileptic patients (Fig.~\ref{fig:ccep}b). By aggregating data from a large cohort of 613 patients—representing stimulation sites across different brain regions — they derived a comprehensive CCEP connectivity matrix of the human brain. This group-level CCEP matrix maps the propagation of neural signals across the cerebral cortex, providing a direct measurement of neural connectivity that is well-suited for validating NPI-inferred EC.}

\re{We compared the NPI-inferred whole-brain EC with the CCEP-derived connectivity matrix (Fig.~\ref{fig:ccep}c). The analysis revealed a significant correlation between NPI-inferred EC and CCEP (left hemisphere, $r = 0.33$, $p < 10^{-3}$), notably higher than the correlation between FC and CCEP (left hemisphere, $r = 0.20$, $p < 10^{-3}$)(Fig.~\ref{fig:ccep}d). Our finding demonstrates that EC inferred from resting-state fMRI data by NPI accurately reflects real neurostimulation propagation pathways and, by extension, the underlying causal relationships between brain regions.}

\re{To illustrate the potential of NPI-inferred EC in guiding neurostimulation, we examined both output and input EC in the CCEP and NPI-inferred EBC matrices (Fig.~\ref{fig:ccep}e). Output EC, represented by a row in the EBC matrix, reflects the propagation range following the stimulation to a specific brain region (i.e., the source). In contrast, input EC, represented by a column in the EBC matrix, indicates the regions capable of propagating stimulation to a given area (i.e., the target). In Fig.~\ref{fig:ccep}f,g, we focused on the output EC using the dlPFC as the source and the input EC using the PCC as the target, as these regions are commonly utilized in neuromodulation studies. The results demonstrate that NPI-inferred EBC accurately captures both output and input patterns, with stronger correlations to CCEP-derived output and input connectivity compared to FC.}

\re{Notably, the advantages of NPI-inferred EBC go beyond those of CCEP. While CCEP-derived EBC relies on invasive procedures involving electrical stimulation at a single site per patient, requiring data aggregation across many individuals to create a group-level connectivity map, NPI is a non-invasive, data-driven approach that does not require real stimulation but virtually perturb the surrogate brain. This makes NPI not only easier to implement but also more adaptable for widespread research and clinical applications. Its non-invasiveness allows for subject-specific analysis, enabling personalized medical insights—an advantage that traditional CCEP methods, constrained by their invasive nature, cannot provide.}

\re{To explore the potential of NPI-inferred subject-level EC as a biomarker, we applied the NPI to fMRI data from the Autism Brain Imaging Data Exchange (ABIDE) dataset~\cite{di2014autism} and the Alzheimer’s Disease Neuroimaging Initiative (ADNI) dataset~\cite{petersen2010alzheimer} (Supplementary Note 6, Supplementary Fig. 12). We found that EC performed comparably to FC in classifying healthy individuals versus patients with disease, suggesting that NPI-inferred EC could serve as a viable alternative to FC as a biomarker for brain disorders. Moreover, the directionality inherent to EC provides valuable insights, potentially guiding personalized treatment strategies.}

\section{Discussion}\label{sec12}

NPI is a data-driven framework that maps the whole-brain EC (Fig. \ref{fig:framework}). We applied NPI to rs-fMRI data, elucidating the directionality, strength, and excitatory/inhibitory properties of the large-scale causal relationships in the human brain (Fig.~\ref{fig:ebc}). NPI advances our comprehension of the brain's functional architecture and has the potential to offer insights into the neural underpinnings of cognitive processes~\cite{Xu2015Effective, Mejias2016Feedforward}. To validate the effectiveness of NPI, it was applied to synthetic datasets, where it accurately and robustly revealed the ground-truth EC (Fig.~\ref{fig:validation}). Further applications to rs-fMRI data demonstrate that NPI can reliably uncover brain-wide EC, which is stable across datasets and atlases with different numbers of brain regions (Fig. \ref{fig:robustness}). We published the group-level EBC with various atlases for all to use.

The concept of EC is pivotal in neuroscience but is interpreted differently across methodologies~\cite{Friston2014DCM, Barnett2014MVGC, Singh2020Estimation}. For example, GC views EC as the predictive influence of one brain region over another, while DCM defines it through coupling coefficients within a state-space model. NPI adopts a `perturb and record' approach that aligns with the statistical notion of causality: a perturbation in one variable that significantly alters another indicates a causal link~\cite{Pearl2009Causality, Woodward2016Causation}. Such a definition is \re{congruent} with empirical methods such as optogenetics, where direct regional perturbations are applied and the resultant neural responses are observed to confirm causal interactions~\cite{Kim2023Wholebrain, Hollunder2024Mapping, Bernal-Casas2017Studying}. 

NPI offers several distinct advantages over traditional methodologies of deriving EC. Firstly, NPI enables non-invasive mapping of EC, a stark contrast to conventional approaches that often require invasive procedures, thereby reducing potential risks and expanding the applicability to a broader range of subjects~\cite{Hollunder2024Mapping}. \re{Secondly, compared to other computational approaches, NPI uses ANNs to learn the complex, nonlinear dynamics of brain activity directly from data. This approach does not rely on predefined model structures or assumptions about neural mechanisms, allowing NPI to effectively handle various data types and dynamics that traditional parametric models may fail to capture~\cite{Das2020Systematic}. The flexibility of the ANN model within NPI facilitates the use of advanced machine learning techniques, such as pre-training for constructing group-level surrogate models and fine-tuning for developing individual-level models~\cite{yuan2024brant,Liang2022Online}. Lastly, NPI's versatility extends to its ability to accommodate various forms and scales of perturbations, once the surrogate model is adequately trained. This adaptability, combined with the efficiency of ANNs in processing large fMRI datasets featuring numerous brain nodes, significantly enhances the practicality of NPI across different experimental settings.}

This study employs the NPI technique primarily within the context of rs-fMRI data, using simple impulse perturbations. However, the versatility of the NPI framework extends well beyond this initial application. By customizing ANN architectures and the virtual perturbation protocol, NPI can be adapted to a wide range of neuroimaging modalities, each characterized by unique spatiotemporal features (Supplementary Note 3, Supplementary Fig. 13, 14). The potential applications of NPI are vast, ranging from analyzing the activity of individual neurons to interpreting population-level neural dynamics and large-scale neuroimaging outputs such as EEG and fMRI. The ability of NPI to integrate EC findings across these diverse scales not only deepens our understanding of the brain's structural-functional interplay but also holds the potential to unveil the neural underpinnings of complex cognitive processes.

NPI holds significant promise for therapeutic applications. Firstly, EC maps inferred by NPI have the potential to serve as biomarkers for neurological disorders, aiding in the mechanistic understanding of these conditions by comparing EC patterns between patients and healthy controls. Furthermore, NPI enhances the precision of neurostimulation therapies used in treating conditions such as Parkinson's disease and depression, by providing personalized EC maps~\cite{Schuepbach2013Neurostimulation, Scangos2021Closedloop}. While direct stimulation of deep brain regions is often desired, practical and ethical considerations frequently necessitate targeting more accessible cortical areas. Thus, understanding the pathways of stimulation propagation within the brain is crucial for selecting optimal control nodes for neurostimulation. \re{To validate this, we compared NPI-inferred EC with actual stimulation propagation matrices obtained through CCEP. Results indicate that NPI-inferred EC mirrors the group-level CCEP patterns, suggesting its utility in guiding personalized neurostimulation strategies (Fig.~\ref{fig:ccep})}. Additionally, NPI's capability to model the effects of stimulating multiple regions or varying stimulation parameters provides a robust framework for optimizing neurostimulation strategies, potentially improving therapeutic outcomes by customizing interventions to individual brain connectivity profiles.

The NPI framework is a data-driven approach that leverages the predictive capabilities of ANNs to infer EC. It inherits a major challenge of data-driven approaches, that is, the necessity for considerable volumes of high-quality data. A pivotal future direction involves developing surrogate brain models that maintain high predictive accuracy without the need for extensive data. This could include exploring advanced ANN architectures that are effective with shorter neural signals or integrating domain-specific knowledge to enhance model performance. Beyond merely inferring EC, another promising avenue is to apply varied interventions to the trained surrogate ANN model, which may deepen our understanding of the real brain's dynamics and potentially uncover new insights into brain function.

\newpage

\section{Methods}

\subsection{The NPI method}

\subsubsection{Training artificial neural network as a surrogate brain}

The ANN in NPI is designed to model the brain's neural dynamics. It can be implemented using various network architectures. \re{In this study, we employ a multi-layer perceptron (MLP) as the surrogate ANN $f(\cdot)$, which predicts the neural state at the next time step based on the states from the three preceding steps} (see Supplementary Note 2 for an optional 1-step input ANN model). The brain's dynamical system is modeled as

\re{\begin{equation}
    \hat{\mathbf{x}}_{t+1} = f(\mathbf{x}_{t}, \mathbf{x}_{t-1}, \mathbf{x}_{t-2},\theta)
\end{equation}}

Here, $\mathbf{x}_{t}$, $\mathbf{x}_{t-1}$, and $\mathbf{x}_{t-2}$ are vectors representing the neural states of various brain regions at times $t$, $t-1$, and $t-2$ respectively. The function $f$ is the MLP model with \re{parameters} $\theta$, which includes all trainable weights of the network. $\hat{\mathbf{x}}_{t+1}$ denotes the MLP-predicted neural state at $t+1$. \re{The network comprises an input layer sized at $3N$, two hidden layers sized at $2N$ and $0.8N$ respectively, and an output layer sized at $N$ for a dataset involving $N$ regions. The network structure is tailored based on the prediction performance on the test set, optimized by grid search (Extended Data Fig. 1).}

\re{The MLP is trained by minimizing the one-step-ahead prediction error. Each training sample contains input $\mathbf{x}_{t}$, $\mathbf{x}_{t-1}$, and $\mathbf{x}_{t-2}$ and output $\mathbf{x}_{t+1}$. The loss function $\mathcal{L}(\theta)$ is formulated as the prediction error between the MLP’s output and the actual next neural state $\mathbf{x}_{t+1}$}

\re{\begin{equation}
    \mathcal{L}(\theta)=\|f(\mathbf{x}_t, \mathbf{x}_{t-1}, \mathbf{x}_{t-2},\theta)-\mathbf{x}_{t+1}\|_2^2
\end{equation}}

\re{Training is conducted over 60 epochs with a batch size of 100, using the Adam optimizer at a learning rate of $10^{-3}$. Implementation was in \textit{PyTorch} on an NVIDIA GeForce RTX 4080 GPU.}

\subsubsection{Perturbing the trained ANN to infer EC}

After training, we perturb each input node of the ANN sequentially to infer whole-brain EC. A perturbation involves a selective increase in the signal of one specific region at time $t$ while keeping other regions unperturbed. EC from region $i$ to all others is quantified as the averaged response at time $t+1$ after applying perturbation to region $i$ at time $t$:

\re{\begin{equation}
    \text{EC}_{i \boldsymbol{\cdot}}=\mathbb{E}_{t}[f(\mathbf{x}_{t}+\Delta\cdot\boldsymbol{e}_i, \mathbf{x}_{t-1}, \mathbf{x}_{t-2})-f(\mathbf{x}_{t}, \mathbf{x}_{t-1}, \mathbf{x}_{t-2})],
\end{equation}}

Here, $\mathbf{e}_i$ is a unit vector with a value of 1 at the $i^{th}$ entry and 0 elsewhere, representing a perturbation in the $i^{th}$ region. $\Delta$ represents the strength of the perturbation, set at half the standard deviation of the BOLD signals. \re{Given} the nonlinear nature of brain dynamics, the response to perturbation varies with brain states, similar to the state-dependent responses observed in real stimulation~\cite{Scangos2021Statedependent, Lurie2020Questions}. To account for this, we conducted virtual perturbation experiments at each time point's state. The subject-level EC was obtained by averaging the responses across all states. Group-level EC and FC were derived by numerically averaging connection strengths across subjects.

\subsection{Ground-truth neural dynamical models for synthetic data generation and NPI validation}

We validated the performance of NPI using a public synthetic fMRI dataset and two generative models with known ground-truth EC including an RNN model and a whole-brain model (WBM). In simulated models, ground-truth EC was obtained by perturbing the activities of a node and observing the propagation of the perturbation among other nodes.

\subsubsection{Synthetic data generated by ground-truth RNN models}

RNN is designed with $n$ nodes. We denote the state of the $i^{th}$ neuron as $x_i$ and $\mathbf{x} = [x_1,...x_n]^T$ is a $n$-dimensional vector that represents the states of all the $n$ neurons in the network. The dynamics of $\mathbf{x}$ are given by the following equation:

\begin{equation}
{\rm d}\mathbf{x}(t)=[-\mathbf{x}(t)+\mathbf{W}\cdot h(\mathbf{x}(t))]\cdot{\rm d}t+\sigma\cdot{\rm d}\mathbf{\xi}(t),
\end{equation}
where $\mathbf{W}$ is the weight matrix, which is defined as SC, and $h()$ is the $\tanh$ activation function. The entries of the weight matrix $\mathbf{W}$ are independent identically distributed centered Gaussians $\mathcal{N}(0, n^{-1/2})$. The initial state is sampled from a Gaussian distribution $\mathcal{N}(0, 1)$. The $\sigma$ is the scaling factor of the Gaussian white noise ${\rm d}\mathbf{\xi}(t)$ with variance $\mathbf{I}_n$. The RNN dynamics are simulated with the Euler method where $\Delta t=0.01$: 

\begin{equation}
\mathbf{x}(t+\Delta t) = \mathbf{x}(t) + [-\mathbf{x}(t)+\mathbf{W}\cdot h(\mathbf{x}(t))]\cdot\Delta t + \sigma\sqrt{\Delta t}\cdot\mathbf{Z}(t),\,
\mathbf{Z}(t)\sim\mathcal{N}(\mathbf{0},\,\mathbf{I}_n) .
\end{equation}

We extracted the dynamics of $\mathbf{x}$ with TR=1 (take 1 point for every 100 pints) to be the training data of NPI. 

The ground-truth EC of RNN is obtained by perturbing the neural states at time $t$ and observing the perturbation-induced response at time $t+1$. To get the ground-truth EC from node $i$ to all other nodes, we perturb the initial signal from $\mathbf{x}_t$ to $\mathbf{x}_t+\Delta\cdot\boldsymbol{e}_i$ with $\Delta=1$. Then we run RNN (100 times for $\Delta t=0.01$) to get $\mathbf{x}_{t+1}$. The ground-truth EC is obtained as the difference between $\mathbf{x}_{t+1}$ mapped from perturbed $\mathbf{x}_{t}$ and unperturbed $\mathbf{x}_{t}$.

\subsubsection{Public synthetic BOLD dataset with few brain regions}

The data generation process involves neural firing rate dynamics followed by a hemodynamic response function (HRF) that transforms the neural signals into BOLD signals. The SC follows a specific topology where most values are 0, with only a few selected positions having non-zero values sampled from a Gaussian distribution with mean = 0.5, standard deviation = 0.1, and values truncated between 0.3 and 0.7. The dataset encompasses 9 network structures with varying degrees of complexity, all of which feature cyclic structures. The number of nodes in these networks ranges from 5 to 10, considering different structures such as unidirectional connections, 2-cycles, and 4-cycles. In the simulation process, the temporal evolution of the neural firing rate follows the linear approximation:

\begin{equation}
\frac{{\rm d}z}{{\rm d}t}=\sigma\mathbf{A}z+\mathbf{C}u,
\end{equation}
where $z$ is a vector representing the firing rate of the regions of interest, $\sigma$ is a constant that controls the neuronal lag within and between nodes, $A$ is the SC matrix between nodes, and $C$ is a matrix that measures the impact of external inputs on the network. The observed BOLD signals are obtained by passing the firing rate $z$ through a hemodynamic response function:

\begin{equation}
    \tilde{y}=g(z,\theta),
\end{equation}
where $\tilde{y}$ is a vector of observed BOLD signals; $g$ is the applied hemodynamic response function; and $\theta$ is a vector of parameters of the function.

From this open dataset, we do not have access to the ground-truth EC. We thus measured the performance of EC inference as the Area Under the Receiver Operating Characteristic Curve (AUC) of classifying the presence or absence of each possible SC connection after binarizing inferred EC.

\subsubsection{Synthetic BOLD data using a whole-brain model (WBM)}

The dynamic mean field model, proposed by Deco et al., is a computational framework that incorporates realistic biophysical properties of neurons and synapses and aims to describe the large-scale dynamics of the human brain~\cite{Deco2013RestingState}.

Consider $N=66$ excitatory neural assemblies with recurrent self-coupling $w=0.55$ and long-range excitatory coupling $G=3.5$. Let $r_i$ and $I_i$ be the population-firing rate and total synaptic input current for population $i\in\{1,\ldots,N\}$. The firing rate $r_i$ is determined by the transfer function $F(I_i)$ given by:

\begin{equation}
r_i = F(I_i) = \frac{aI_i - b}{1-\exp\left(-d\left(aI_i-b\right)\right)}
\end{equation}
where $a$ = 270 Hz/nA, $b$ = 108 Hz, $d$ = 0.154 sec. The net current $I_i$ into population $i$ is given by 

\begin{equation}
I_i = w J_N S_i + G J_N\sum_{j=1}^N C_{ij} S_j + I_{bi}
\end{equation}
where $J_N = 0.2609$ is the overall excitatory strength. The coupling parameters $w$ and $G$ scale the strengths of local and long-range interactions, respectively. Structural connectivity $C$ is extracted from healthy humans using diffusion spectrum imaging (DSI)~\cite{hagmann2008mapping}. $I_{bi}$ is the background input into population $i$, which has a mean ($I_0$) and a noise component described by an Ornstein-Uhlenbeck (OU) process: 

\begin{equation}
\tau_0\frac{{\rm d}I_{bi}}{{\rm d}t} =-(I_{bi}-I_0)+\eta_i(t)\sqrt{\tau_0\sigma^2}
\end{equation}
where $I_0=0.3255$ nA, filter time constant $\tau_0=2$ ms, and noise amplitude $\sigma=0.02$ nA; $\eta(t)$ is a Gaussian white noise which has zero means with standard deviation equals one.
Assume that synaptic drive variable $S_i$ for population $i$ obeys:

\begin{equation}
\frac{{\rm d}S_i}{{\rm d}t} = F\left(I_i\right)\gamma\left(1-S_i\right)-\frac{1}{\tau_s}\,S_i
\end{equation}
where synaptic time constant $\tau_s$ = 100 ms and $\gamma$ = 0.641.
The synaptic drive $S_i$ is indicative of the level of activity in population $i$ at time $t$. The BOLD signal $B_i(t)$ is typically modeled as a delayed low-pass filtered version of $S_i(t)$. We use the Boynton gamma function as the filter kernel~\cite{boynton1996linear}:

\begin{equation}
f_{bold}(t)=\left(\frac{t-o}{\tau_{bold}}\right)^{p-1}\,\frac{1}{(p-1)!}\,\exp\left(-\frac{t-o}{\tau_{bold}}\right) H(t-o)
\end{equation}
where $p=2$ is a shape parameter, $\tau_{bold}=1.25$ s is a timescale parameter and $o=2.25$ s is a delay parameter and $H(t-o)$ is the Heaviside function.
The BOLD signal $B(t)$ generated by $S_i(t)$ is computed by evaluating the convolution of $S_i(t)$ with filter kernel $f_{bold}(t)$:

\begin{equation}
B_i(t) = \int_{-\infty}^t S_i(x)\,f_{bold}(t-x)\,{\rm d}x
\end{equation}

We extracted the dynamics of BOLD signals with TR=0.72 (the same as HCP data) to be the training data of NPI. 

To derive the ground-truth EC of WBM, we perturb the total synaptic input current $I$ at time $t$. Due to the time lag in HRF, the perturbation-induced response starts to be observed at time $t+4$ TR. To get the ground-truth EC from node $i$ to all other nodes, we perturb the initial signal from $I_t$ to $I_t+\Delta\cdot\boldsymbol{e}_i$ with $\Delta=5$. Then we simulate the WBM and get the BOLD signals at time $t+4$ TR. The ground-truth EC is obtained as the difference between BOLD signals at time $t+4$ TR mapped from perturbed $I_{t}$ and unperturbed $I_{t}$.

\subsection{Data processing}

\re{In this study, we used real data from HCP dataset for healthy subjects~\cite{VanEssen2013WUMinn}, ABCD dataset for healthy subjects~\cite{saragosa2022practical}, CCEP dataset for patients with epilepsy~\cite{ccep}, ABIDE dataset for patients with autism~\cite{di2014autism} and ADNI dataset for Alzheimer's disease~\cite{petersen2010alzheimer}. Specifically,} for the HCP dataset, we used resting-state fMRI (rs-fMRI) data from 800 healthy subjects from the HCP S1200 release~\cite{VanEssen2013WUMinn}. The rs-fMRI data were recorded with a TR of 0.72 seconds, with each subject undergoing four 15-minute sessions. The data were then preprocessed using multi-modal inter-subject registration (MSMAll)~\cite{Robinson2014MSM}. For the ABCD dataset, we used rs-fMRI data from 2000 healthy subjects, also recorded with a TR of 0.72 seconds.

The rs-fMRI data from the HCP and ABCD datasets were preprocessed using the HCP minimal preprocessing pipeline~\cite{GLASSER2013Minimal}. Denoising was performed with ICA-FIX, which removes structured noise by combining independent component analysis with the FSL tool \textit{FIX}. The denoised data were then further processed using the \textit{Nilearn} package~\cite{Abraham2014Machine} to extract regional-level BOLD signals in the 0.01 to 0.1 Hz frequency range.

\re{When evaluating the signal prediction performance of the surrogate models, each model is trained on 90\% of the individual’s fMRI data (i.e., the full first three sessions and 60\% of the fourth session) and tested on the remaining 10\% (i.e., the final 40\% of the fourth session). On the other hand, FC is estimated by calculating Pearson’s correlation coefficient between the time series of each pair of brain regions, using data from all four sessions.
Similarly, when applying NPI to infer the individual EC, all four sessions are used for training the surrogate model.}

To analyze the similarity between the structural connectivity (SC) and EC, we used the SC matrix constructed by Demirtaş et al.~\cite{Demirtas2019Hierarchical}, derived using FSL's \textit{bedpostx} and \textit{probtrackx2} workflows, which count the number of streamlines intersecting white and gray matter. The SC matrix is scaled to a range of $0$ to $1$ and then log-transformed. The EC matrix for each subject is obtained from the NPI framework, which is trained on four fMRI runs per subject. The EC is then averaged across 800 subjects and scaled so that the strongest connection has a value of one.

In the analysis of the HCP and ABCD datasets, the brain is parcellated into 379 regions according to the Multi-Modal Parcellation (MMP 1.0) atlas~\cite{Glasser2016Multimodal}, which includes 180 cortical regions in each hemisphere and 19 subcortical regions. The analysis focuses on the EC among the 360 cortical regions, with subcortical regions incorporated during training to reduce bias in EC inference from unobserved regions. Parcellation is conducted by averaging BOLD signals across voxels within each cortical region.

The parcellated 360 cortical regions are assigned to seven functional networks, according to the resting-state networks defined in Yeo et al. \cite{ThomasYeo2011organization}. The seven functional networks are visual network (VIS), somatomotor network (SOM), dorsal attention network (DAN), ventral attention network (VAN), \re{limbic network (LIM)}, frontoparietal control network (FPN), and default mode network (DMN). \re{Each region is assigned to the functional network with which it shares the most voxels.} We place the seed region in the left-hemisphere core brain region of each of the seven functional networks (seeds are shown in Supplementary Table S1). Then we calculate the seed-based FC using Pearson's correlation between the seed region and \re{all other regions}.

\re{The cortico-cortical evoked potentials (CCEP) data are provided by the F-TRACT atlas with MMP parcellation~\cite{ccep}. For the comparison, we use the EBC matrix that NPI inferred from the HCP rs-fMRI data using the same atlas as CCEP. The detailed description of data analysis for the ABIDE and ADNI datasets is in Supplementary Note 6.}


\subsection{Quantitative metrics and statistics}

To measure the goodness of brain signal prediction, we calculated the coefficient of determination ($R^2$) between the ground-truth signal and predicted signal for each brain region, using the following formula, $R^2 = 1 - \frac{\sum_{i=1}^{n} (y_i - \hat{y}_i)^2}{\sum_{i=1}^{n} (y_i - \bar{y})^2} $, where $y_i$ represents the actual signals, $\hat{y}_i$ represents the predicted signals, $\bar{y}$ is the mean of the actual signals, $n$ is the number of time points. Overall $R^2$ is the averaged $R^2$ across all brain regions. 

To assess ANN's ability to learn inter-regional relations, we calculated Pearson's correlation coefficient ($r$) between model FC and empirical FC. Model FC was obtained by the model-generated data with 1200 TRs, where we recurrently fed ANN's output as input to generate BOLD signals. The empirical FC was \re{obtained} by calculating the inter-region correlation coefficient of the ground-truth data. 

To assess the performance of EC inference, we calculated Pearson's correlation coefficient
$r$ between ground-truth EC and NPI-inferred EC. For matrices with binary weights (Fig.~\ref{fig:validation}i,j), we calculated the Area Under the Receiver Operating Characteristic Curve (AUC) to assess the model's ability to distinguish the presence or absence of specific connections correctly.

\section*{Acknowledgments}

This work is supported by the National Key R\&D Program of China (2021YFF1200804), Shenzhen Excellent Youth Project (RCYX20231211090405003), Shenzhen Science and Technology Innovation Committee (2022410129, KJZD20230923115221044, KCXFZ20201221173400001), Guangdong Provincial Key Laboratory of Advanced Biomaterials (2022B1212010003), Hong Kong RGC Senior Research Fellowship Scheme (SRFS2324-2S05).
We thank professors Haiyan Wu, Jing Jiang, Kai Du, Yu Mu, Pengcheng Zhou, Shi Gu, Zaixu Cui, and members of the NCC lab including Chen Wei, Kexin Lou, Zongsheng Li, Xin Xu, and Song Wang for helpful discussions and reviewers for their insightful suggestions. 

\section*{Declarations}

All the authors declare no conflict of interest.

\section*{Data availability}

Our synthetic data (generated by a ground-truth RNN and a whole-brain model) are publicly available at \url{https://github.com/ncclab-sustech/NPI/}. The HCP dataset is available at \url{https://www.humanconnectome.org/study/hcp-young-adult/document/1200-subjects-data-release}. The ABCD dataset is available at \url{https://abcdstudy.org/scientists/data-sharing/}. The ABIDE dataset is available at \url{http://fcon_1000.projects.nitrc.org/indi/abide/}. The CCEP dataset is available at \url{https://f-tract.eu/atlas/.}

\section*{Code availability}
Codes for using NPI are available at \url{https://github.com/ncclab-sustech/NPI/}.

\newpage

\bibliography{sn-bibliography.bib}

\newpage
\section{Extended Data}

\renewcommand{\figurename}{Extended Data Fig.}
\setcounter{figure}{0}

\renewcommand{\tablename}{Extended Data Table}
\setcounter{table}{0}

\begin{figure}[H]
\centering
\includegraphics[width=0.6\linewidth]{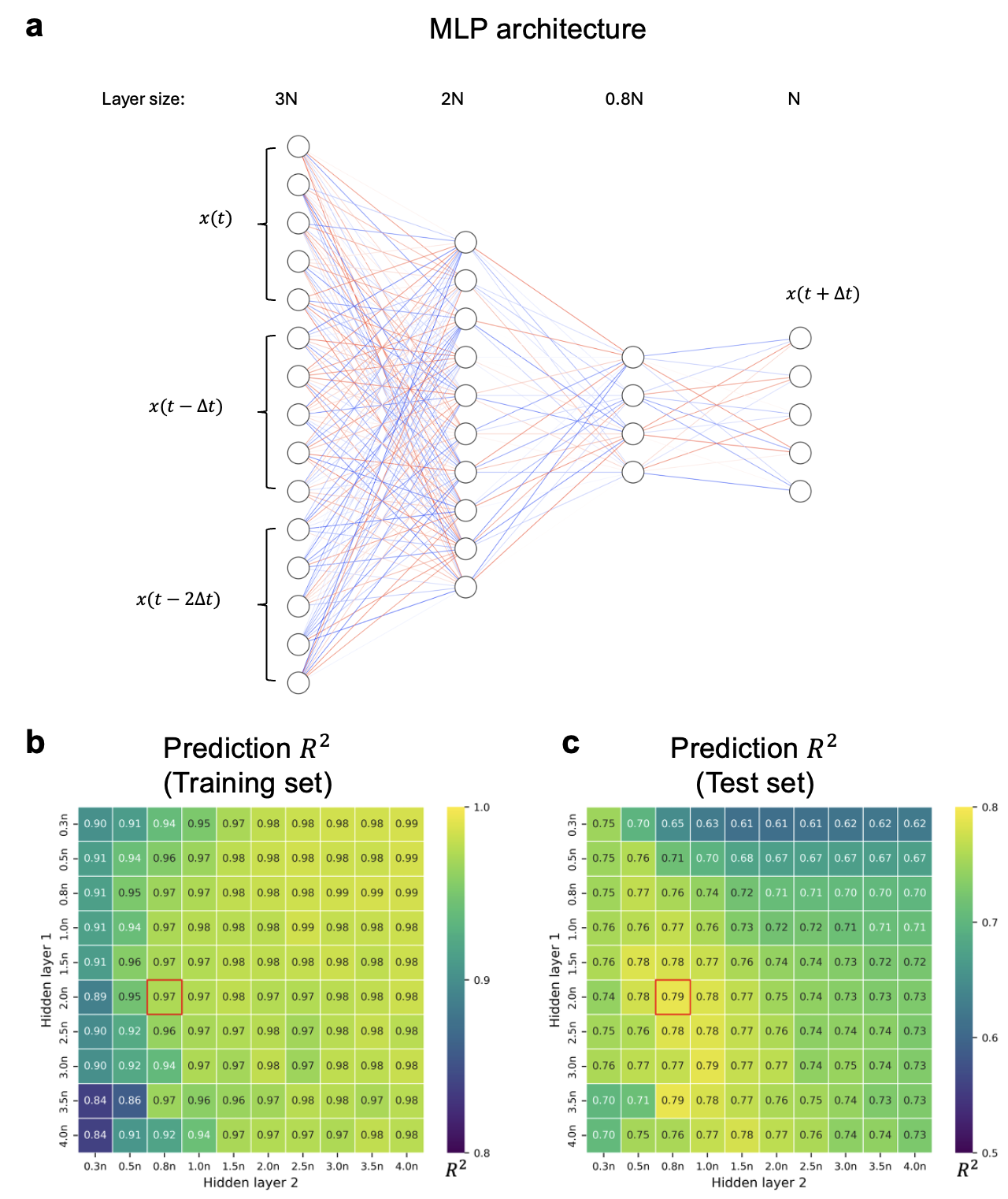}
\caption{\re{\textbf{Optimizing MLP architecture via grid search.} 
\textbf{a}, The MLP architecture used in our study, which includes an input layer, two hidden layers, and an output layer, derived from grid search.
\textbf{b}, The $R^2$ of one-step-ahead prediction of the training set under various sizes of hidden layer configurations, averaged across 20 subjects. 
\textbf{c}, The $R^2$ of one-step-ahead prediction of the test set under various sizes of hidden layer configurations, averaged across 20 subjects.}
}
\label{fig:MLP}
\end{figure}

\newpage
\begin{figure}[H]
\centering
\includegraphics[width=0.9\linewidth]{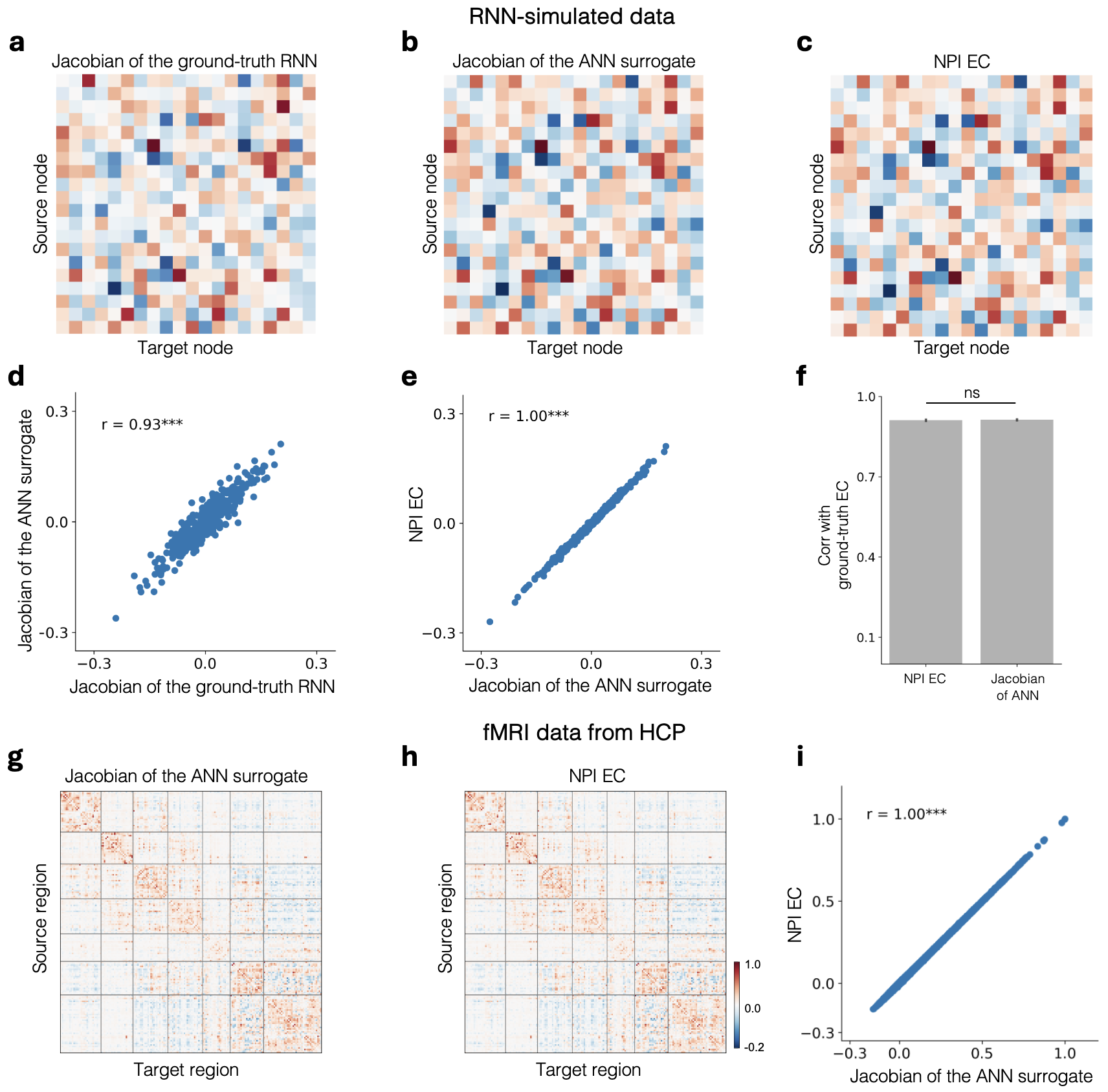}
\caption{\re{\textbf{The EC inferred by NPI is consistent with the Jacobian matrix of the trained ANN model}.
\textbf{a}, Jacobian matrix of an example RNN, numerically calculated using Pytorch.
\textbf{b}, Jacobian matrix of an ANN trained to predict the synthetic signal generated by the RNN.
\textbf{c}, NPI-inferred EC by perturbing the trained ANN.
\textbf{d}, Jacobian of the trained ANN vs. Jacobian matrix of the ground-truth RNN across connection pairs.
\textbf{e}, NPI-inferred EC vs. Jacobian matrix of trained ANN across connection pairs.
\textbf{f}, Correlation coefficients between the NPI-inferred EC and the ground-truth EC, and between the Jacobian matrix and the ground-truth EC.
\textbf{g}, NPI-inferred EC on resting-state fMRI data from the HCP dataset, averaged across 800 subjects.
\textbf{h}, Jacobian matrix of the ANN model trained on resting-state fMRI data from the HCP dataset, averaged across 800 subjects.
\textbf{i}, NPI-inferred EC vs. Jacobian matrix of the trained ANN across connection pairs.
}}
\label{fig:jacobian}
\end{figure}

\newpage
\begin{figure}[H]
\centering
\includegraphics[width=1\linewidth]{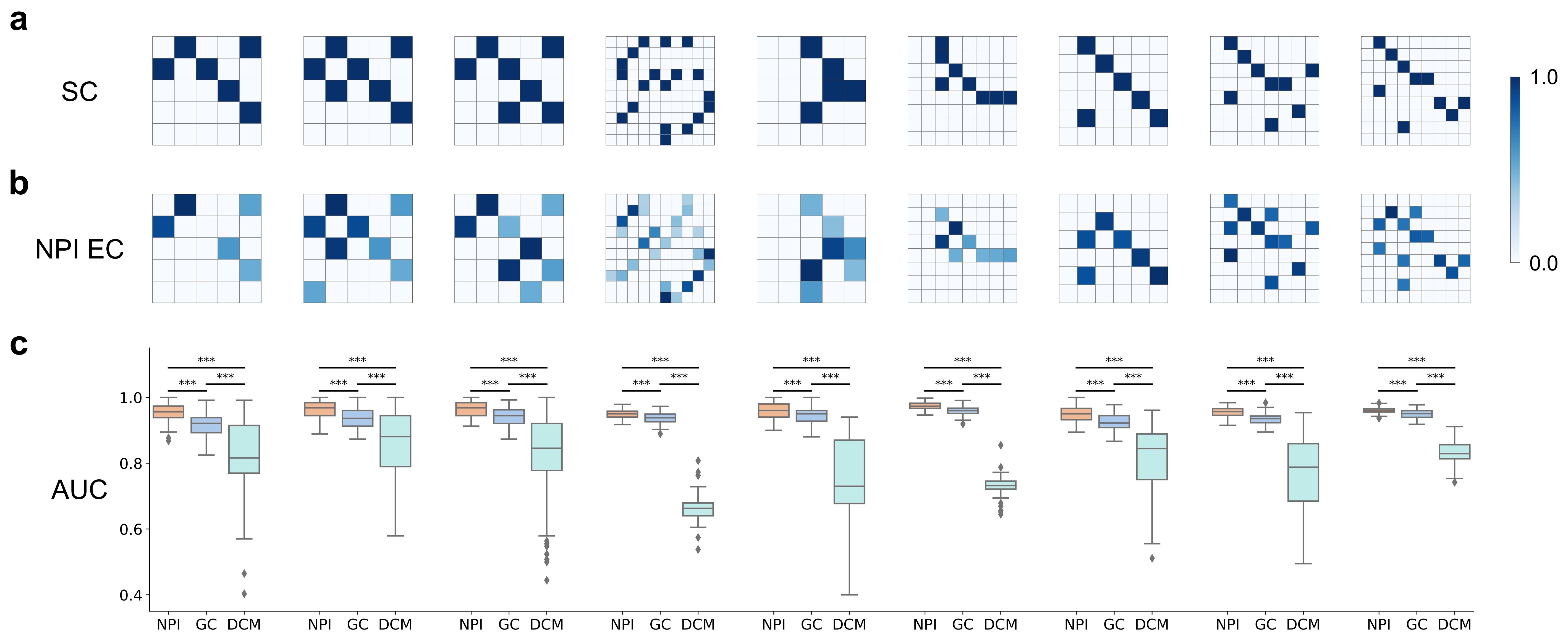}
\caption{\textbf{The performance of NPI is reliable across different network \re{topographies}}. 
\textbf{a}, The test data are generated by generative models with pre-defined directed, binary structural connectivities (SC), from a public dataset proposed by Sanchez-Romero et al.~\cite{Sanchez-Romero2019Estimating}. The weights in this linear generative model are set to either 0 or 1, delineating the presence or absence of direct SC. The neural firing rates generated by this model are subsequently converted into BOLD signals through the application of a hemodynamic response function (HRF).
\textbf{b}, NPI is utilized to map the EC from these BOLD signals, mapping the causal neural interactions. 
\textbf{c}, Comparisons of the AUC scores for EC inference across nine different SC configurations show the superior performance of NPI over GC and DCM. Error bars represent standard deviation.
}
\label{fig:opendata}
\end{figure}

\newpage
\begin{figure}[H]
\centering
\includegraphics[width=1\linewidth]{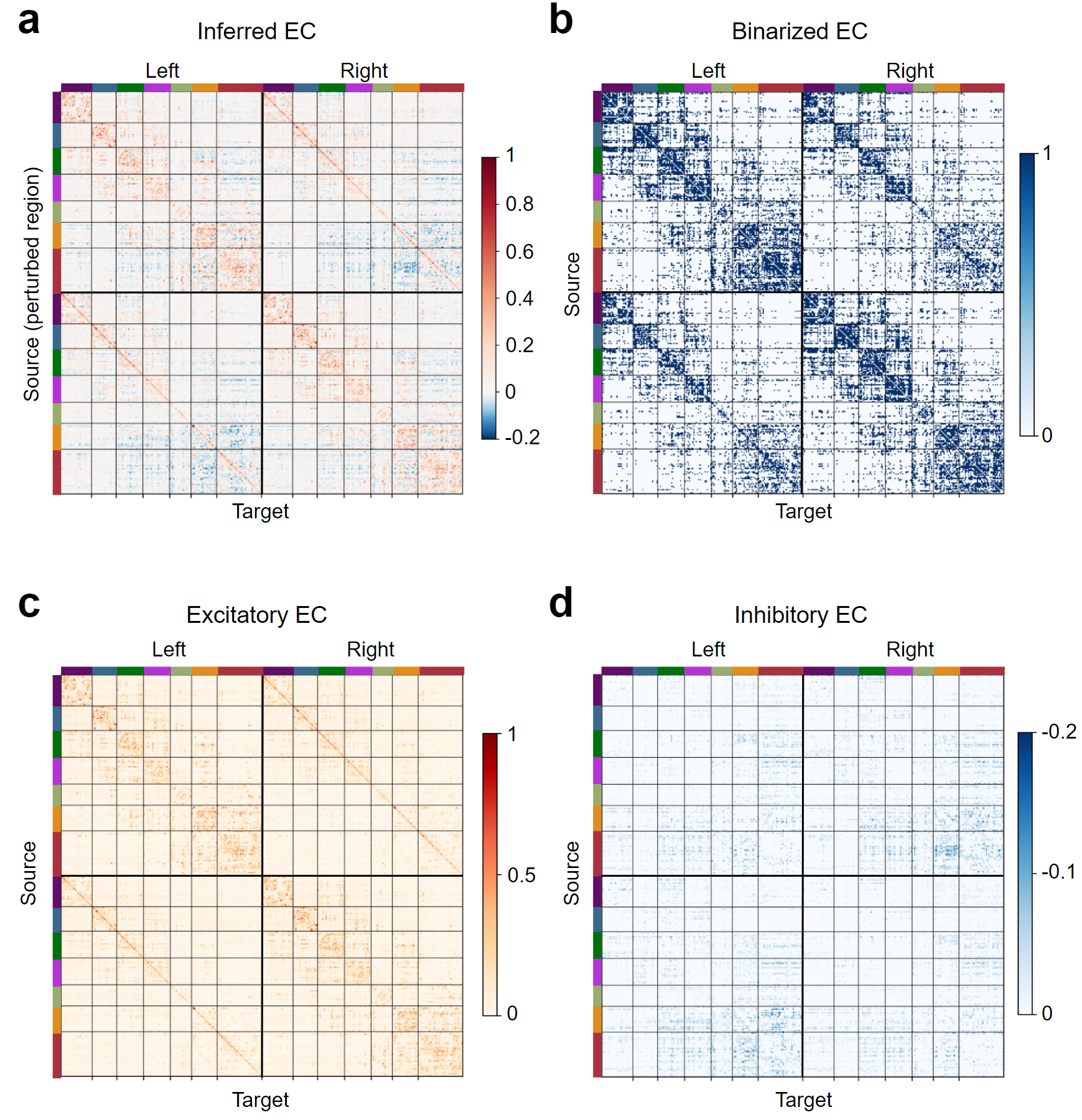}
\caption{\textbf{The EBC, binarized EBC, and excitatory and inhibitory part of EBC}. 
\textbf{a}, The whole-brain EBC.
\textbf{b}, EBC binarized by a threshold larger than 80\% of EC. The entries larger than the threshold are set to 1, while the rest are set to 0. 
\textbf{c,d}, The excitatory (c) and inhibitory (d) parts of EBC.
}
\label{fig:wholeec}
\end{figure}

\newpage
\begin{figure}[H]
\centering
\includegraphics[width=1\linewidth]{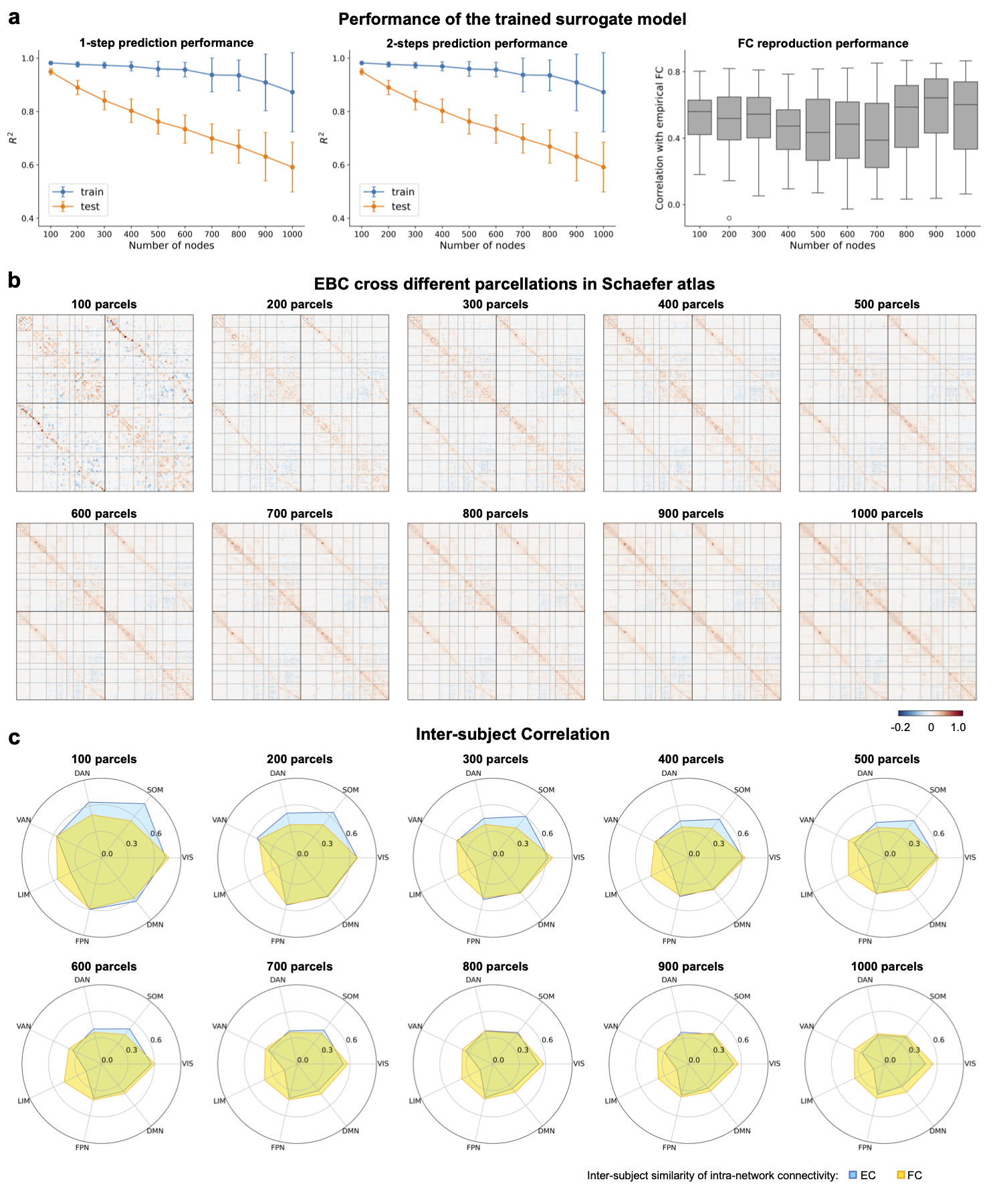}
\caption{\re{\textbf{The NPI-inferred EC derived from brain atlases with increasing numbers of regions.}
NPI was applied to BOLD signal parcellated from Schaefer atlas with 100, 200, ..., up to 1000 brain regions~\cite{schaefer2018local}. The fMRI data from 100 randomly selected individuals from the HCP dataset are used.
\textbf{a}, \textbf{Performance of the trained surrogate model.} Left: 1-step prediction $R^2$ (left) and 2-step prediction $R^2$ (right) on training (blue) and test (orange) sets as the number of brain regions increases. The error bars represent the standard deviation (s.t.d.) calculated across subjects. Right: FC reproduction performance as the number of brain regions increases. For each participant, the correlation between model FC and empirical FC is computed. The central line and bounds represent the median performance and the 25th and 75th percentiles.
\textbf{b}, \textbf{EBC across different parcellations in Schaefer atlases.} The group-level EBC are averaged across 100 subjects.
\textbf{c}, \textbf{Inter-subject correlation of individual EC and FC across different parcellations in Schaefer atlases.} Results are averaged across subjects.}}
\label{fig:atlas1000}
\end{figure}

\newpage

\section*{Supplementary Notes, Figures and Tables}\label{supp}

Supplementary Note 1: Alternative implementations of ANN models. \\
Supplementary Note 2: 1-step input MLP v.s. 3-step input MLP. \\
Supplementary Note 3: Alternative perturbations. \\
Supplementary Note 4: EC inferred as the expected Jacobian matrix of the trained ANN. \\
Supplementary Note 5: Implementation of competing methods. \\
Supplementary Note 6: Applying NPI on ADNI and ABIDE datasets. \\

\noindent
Supplementary Fig. 1: Optional surrogate model architectures.  \\
Supplementary Fig. 2: Signal changes after perturbing a node of RNN. \\
Supplementary Fig. 3: Ground-truth EC and NPI-inferred EC of RNN. \\
Supplementary Fig. 4: NPI outperforms Granger causality (GC) in EC inference from ground-truth RNN generated data. \\
Supplementary Fig. 5: EC from NPI and GC on the synthetic data generated by WBM. \\
Supplementary Fig. 6: NPI-inferred EC on the left hemisphere using WBM-generated BOLD data. \\
Supplementary Fig. 7: Comparison across different prediction models on WBM-generated data. \\
Supplementary Fig. 8: Performance comparison between single-step-input and three-step-input ANN. \\
Supplementary Fig. 9: NPI-inferred EBC from the HCP dataset using the AAL atlas. \\
Supplementary Fig. 10: The EC with the strongest strengths and regions with the largest degrees. \\
Supplementary Fig. 11: $p$-values of EC inferred from the HCP dataset. \\
Supplementary Fig. 12: Applying NPI to the ABIDE dataset and ADNI dataset. \\
Supplementary Fig. 13: EC obtained by incorporating hemodynamic convolution in ANN perturbation. \\
Supplementary Fig. 14: EC obtained by positive/negative impulse perturbation to the surrogate model. \\
Supplementary Fig. 15: Correlation of intra-network FC of seven functional networks across sessions, subjects, and datasets. \\

\noindent
Supplementary Table 1: Performance of one-step-ahead prediction, FC reproduction, and EC inference across surrogate models and datasets. \\
Supplementary Table 2: Correlation between the NPI-inferred EC and the SC across surrogate models and datasets. \\
Supplementary Table 3: MNI Coordinates of seed regions in seven brain functional networks. \\
Supplementary Table 4: Fitting the distribution of EBC strengths to various density functions. \\
Supplementary Table 5: Name and order of the MMP atlas for the left hemisphere. Regions in the right hemisphere are ordered using the same order.

\end{document}


\linenumbers
 

\title[Supplementary materials]{\vspace{-2cm} Mapping effective connectivity by virtually perturbing a surrogate brain}

\author[1]{\fnm{Zixiang} \sur{Luo}}
\author[1]{\fnm{Kaining} \sur{Peng}}
\author[1]{\fnm{Zhichao} \sur{Liang}}
\author[1]{\fnm{Shengyuan} \sur{Cai}}
\author[2]{\fnm{Chenyu} \sur{Xu}}
\author[1]{\fnm{Dan} \sur{Li}}
\author[3]{\fnm{Yu} \sur{Hu}}
\author[4]{\fnm{Changsong} \sur{Zhou}}
\author*[1]{\fnm{Quanying} \sur{Liu}}\email{liuqy@sustech.edu.cn}

\affil[1]{\orgdiv{Department of Biomedical Engineering}, \orgname{Southern University of Science and Technology}, \orgaddress{\city{Shenzhen}, \postcode{518055}, \country{China}}}

\affil[2]{\orgdiv{Department of Electrical and Computer Engineering}, \orgname{Iowa State University}, \orgaddress{\city{Ames}, \postcode{50011}, \state{Iowa}, \country{USA}}}

\affil[3]{\orgdiv{Department of Mathematics and Division of Life Science}, \orgname{The Hong Kong University of Science and Technology}, \orgaddress{\street{Hong Kong SAR}, \country{China}}}

\affil[4]{\orgdiv{Department of Physics}, \orgname{Hong Kong Baptist University}, \orgaddress{\street{Hong Kong}, \country{China}}}

\maketitle

\newpage
\section*{Supplementary materials}\label{supp}

\subsection*{Supplementary Note 1: Alternative implementations of ANN models}\label{supp-note1}

NPI is a flexible framework that can incorporate various artificial neural network (ANN) architectures as surrogate models for the brain. While the results presented in the main text are based on using a Multilayer Perceptron (MLP) as the surrogate model, the MLP can be replaced by other neural network architectures or any time-series prediction models. Here, we implemented four models: MLP, Convolutional Neural Network (CNN), Recurrent Neural Network (RNN), and Vector Autoregression (VAR). We compared the MLP to these alternative models to evaluate their performance in signal prediction, functional connectivity (FC) reproduction, and effective connectivity (EC) inference. Each model was designed to take the preceding three time points of fMRI data as input to predict the subsequent time point.

\subsubsection*{Recurrent neural network}

The RNN model, particularly suited for sequential data prediction, captures recurrent dependencies in time-series data through recurrent units. \re{Here, we consider a generic state-space model with an encoder module, a recurrent neural network, and a decoder module. The encoder consists of two fully connected linear layers and a ReLU activation function between them. The encoder is designed to infer the latent state of the original signal. The recurrent neural network predicts the future latent state given the current latent state. The decoder is a linear layer that maps the predicted latent state to the original state space. A one-step prediction loss is calculated and minimized during the training process. All the modules are jointly trained using gradient descent with Adam optimizer and learning rate $10^{-3}$. The network architecture is summarized as Fig. \ref{figS:RNN-CNN}a.}

\subsubsection*{Convolutional neural network}
\re{The designed convolutional neural network (CNN) adopts a two-layer one-dimensional (1D) convolution, with a Relu activation function between them. Finally, a linear layer maps the output of the last convolution layer to the predicted time series. The network architecture is illustrated in Fig. \ref{figS:RNN-CNN}b. The designed CNN is trained using gradient descent with Adam optimizer and learning rate $10^{-3}$.}

\subsubsection*{Vector auto-regression}

A vector auto-regression model of order $p$ (i.e., VAR($p$)) can be expressed as:
\begin{equation}
\mathbf{y}_t = \mathbf{c} + \sum_{i=1}^p \mathbf{A}_i \mathbf{y}_{t-i} + \mathbf{e}_t,
\end{equation}
where $\mathbf{y}_t$ is a vector of variables at time t. Each variable corresponds to the state of a brain region. $\mathbf{c}$ is a bias vector. $\mathbf{A}_i$ are the coefficient matrices for lag $i$. $\mathbf{e}_t$ is the error terms at time $t$.

The parameters in the VAR(p) model to be estimated are the vector $\mathbf{c}$ and matrices $\mathbf{A}_i$ for $i=1,\ldots,p$. This study used gradient descent with Adam optimizer and learning rate $10^{-3}$ to optimize these parameters. \\

\subsection*{Supplementary Note 2: 1-step input MLP v.s. 3-step input MLP}

The NPI framework is highly flexible and can accommodate various input configurations, whether it is a multi-step-input ANN (as described in the Methods section and Supplementary note 1) or a Markov-like single-step-input ANN $f(\cdot)$, 
\begin{equation*}
    \hat{\mathbf{x}}_{t+1} = f(\mathbf{x}_{t},\theta),
\end{equation*}
where $\theta$ is the vector of all unknown parameters in ANN; $\mathbf{x}_t$ is the vectorized neural data across the cortical areas at time step $t$; $\hat{\mathbf{x}}_{t+1}$ is the predicted neural data at time step $t+1$ by the ANN. 

Take MLP as an example. For a dataset with $N$ regions, the MLP comprises an input layer sized at $N$, two hidden layers sized at $2N$ and $0.8N$ respectively, and an output layer sized at $N$. ANN is trained by minimizing the one-step-ahead prediction error with the preceding \re{one-step} brain states. Each training sample contains input-output pairs with input $\mathbf{x}_{t}$ and output $\mathbf{x}_{t+1}$. The objective function $\mathcal{L}(\theta)$ is formulated as:
\begin{equation*}
    \mathcal{L}(\theta)=\|f(\mathbf{x}_t,\theta)-\mathbf{x}_{t+1}\|_2^2 .
\end{equation*}

In this study, we used the Adam optimizer with learning rate $10^{-3}$ to optimize the network parameters $\theta$ for this 1-step-input MLP, where the batch size is 100 and the number of epochs is 60. 

The one-to-all EC from region $i$ to all other regions is defined as the averaged response at time $t+1$: 
\begin{equation*}
    \text{EC}_{i \boldsymbol{\cdot}}=\mathbb{E}_{t}[f(\mathbf{x}_{t}+\Delta\cdot\boldsymbol{e}_i)-f(\mathbf{x}_{t})],
\end{equation*}
where $\mathbf{e}_i$ is a unit vector in the direction of the $i^{th}$ dimension, which has value $1$ at $i^{th}$ entry and value $0$ elsewhere. $\Delta$ is the strength of perturbation. We set $\Delta$ as half of the standard deviation of the BOLD signals. The final EC is calculated as the averaged response after applying perturbations at all data points used for training. 

We compared the results of the MLP trained with the 1-step-predict-1-step configuration and the 3-step-predict-1-step configuration, including the signal prediction performance (i.e., $R^2$), FC reproduction, and EC inference. Our results underscore that regardless of the input step configuration, the inferred EC remains robust (Supplementary Fig.~\ref{fig:1step}). The inferred EC matrices using two configurations are highly correlated ($r=0.919$, $p<10^{-3}$, Pearson correlation), suggesting that NPI-inferred EC is reliable as long as the surrogate model is well trained to predict future states.

\subsection*{Supplementary Note 3: Alternative perturbations} \label{supp-note2}

\re{Once the surrogate model is trained, perturbations can be introduced to estimate EC, such as applying an impulse perturbation directly to the brain state. In the main text, we applied an impulse perturbation at the time $t$ to the 3-step-input ANN. NPI can be adapted to receive various perturbation patterns, such as an HRF-convolved perturbation.}

\subsubsection*{\re{HRF-convolved perturbation to 3-step-input ANN}}

Considering that BOLD signals in fMRI are influenced by the hemodynamic response function (HRF, $f_{bold}(t)$), external perturbations may affect the brain state in a form convolved with HRF. For instance, a perturbation of $e(t)$ in the brain would lead to a change of $e(t)*f_{bold}(t)$ in the BOLD. 

To do this, we assume that an impulse perturbation with duration 0.02 s and strength 0.05 was applied to underlying neural activities at $t-6$ TR. The resulting HRF $f_{bold}(t)$ is modeled as the convolution of this impulse with a hemodynamic kernel~\cite{boynton1996linear}:
\begin{equation}
f_{bold}(t)=\left(\frac{t-o}{\tau_{bold}}\right)^{p-1}\,\frac{1}{(p-1)!}\,\exp\left(-\frac{t-o}{\tau_{bold}}\right) H(t-o),
\end{equation}
where $p=2$ is a shape parameter, $\tau_{bold}=1.25$ s is a timescale parameter and $o=2.25$ s is a delay parameter and $H(t-o)$ is the Heaviside function. The BOLD signal $B(t)$ generated by $S_i(t)$ is computed by evaluating the causal convolution of $S_i(t)$ with filter kernel $f_{bold}(t)$:
\begin{equation}
B_i(t) = \int_{-\infty}^t S_i(x)\,f_{bold}(t-x)\,{\rm d}x .
\end{equation}

The strongest responses are observed in time $t-2$, $t-1$, and $t$ TR, with strength $\Delta_2$, $\Delta_1$, and $\Delta_0$. We used response values in these three time points as the perturbation applying to the three input steps of the trained three-step-input MLP. The one-to-all EC from region $i$ to all other regions is defined as the averaged response at time $t+1$: 
\begin{equation*}
    \text{EC}_{i \boldsymbol{\cdot}}=\mathbb{E}_{t}[f(\mathbf{x}_{t}+\Delta_0\cdot\boldsymbol{e}_i, \mathbf{x}_{t-1}+\Delta_1\cdot\boldsymbol{e}_i, \mathbf{x}_{t-2}+\Delta_2\cdot\boldsymbol{e}_i)-f(\mathbf{x}_{t}, \mathbf{x}_{t-1}, \mathbf{x}_{t-2})],
\end{equation*}
where $\mathbf{e}_i$ is a unit vector that has value $1$ at $i^{th}$ entry and value $0$ elsewhere. The final EC is calculated as the averaged response after applying perturbations at all data points used for training. 

We applied the HRF-convoluted response to an impulse input as the perturbation to the surrogate model trained with the HCP dataset (Supplementary Fig. 13). The comparison of EC estimates, incorporating hemodynamic convolution against our original NPI method, yielded a high correlation coefficient ($r=0.934$). This underscores the robustness and reliability of NPI. 

\subsection*{Supplementary Note 4: The expected Jacobian matrix of ANN as EC}

One alternative way to derive EC from trained ANN is to calculate the Jacobian matrix of the trained ANN. The Jacobian matrix describes the local behavior of a function that maps from one vector space to another. For a function $f(\cdot)$, the Jacobian matrix consists of all first-order partial derivatives, and it is represented as a matrix $\mathbf{J}$ where each element $\mathbf{J}_{ij}$ is defined as $\frac{\partial y_i}{\partial x_j}$. Here, $y_i$ are the components of the output and $x_j$ are the components of the input vector. It describes how changes in the input vector affect the outputs.

For an MLP that functions as a map $f:\mathbb{R}^n\to\mathbb{R}^m$, where $n$ is the dimension of the input vector and $m$ is the dimension of the output vector, the Jacobian matrix $\mathbf{J}_f$ of $f$ at a certain input vector $\mathbf{x}\in\mathbb{R}^n$ is defined as:
\begin{equation}
    J_f(\mathbf{x})=\begin{bmatrix}\frac{\partial f_1}{\partial x_1}&\cdots&\frac{\partial f_1}{\partial x_n}\\
    \vdots&\ddots&\vdots\\\frac{\partial f_m}{\partial x_1}&\cdots&\frac{\partial f_m}{\partial x_n}\end{bmatrix},
\end{equation}
where $f_i$ are the components of the output vector $f(\mathbf{x})$, and $x_j$ are the components of the input vector $\mathbf{x}$.

To calculate the partial derivative of each output component $f_i$ with respect to each input component $x_j$, backpropagation is employed. It involves using the chain rule to compute gradients of MLP. We used Pytorch function ``\textit{torch.autograd.functional.jacobian}" to automatically compute the Jacobian matrix of trained MLP. 

Like the response to perturbation, the Jacobian matrix also changes with states. We calculate the Jacobian matrix at all states and take the averaged Jacobian matrix $\mathbf{J}$ as the final output. As MLP takes the states at the preceding three time steps as input and predicts the state at the next step, it has input size $3N$ and output size $N$. The size of its Jacobian matrix $\mathbf{J}$ is thus $3N \times N$. Consistent with the regular impulse perturbation method that only applies perturbation to the last time step, we extract the submatrix of $\mathbf{J}$ consisting of the last $N$ rows and all columns of $\mathbf{J}$ as the EC.

To get the Jacobian matrix of RNN, we use the Pytorch function ``torch.autograd.functional.jacobian" to automatically compute the Jacobian matrix of the ground-truth RNN. To have a consistent TR with the RNN-generated data, we compute the Jacobian matrix of a constructed Pytorch function that simulates the discretized ground-truth RNN with $\Delta t=0.01$ for 100 times. 

We compared the expected value of the Jacobian matrix of our trained MLP surrogate model (Extended Data Fig.2). Training on synthetic signals generated by a ground-truth RNN, results show that Jacobian of the trained ANN well captured the Jacobian of the ground-truth RNN. Moreover, no matter the surrogate model is trained with the RNN-simulated data or with the real fMRI data from the HCP dataset, defining EC as the response to perturbing the trained ANN and as the Jacobian matrix of the trained ANN yield similar results ($r = 1.00$, $p < 10^{-3}$, Pearson correlation). This indicates that the perturbation magnitude we use are probably sufficiently small so that the NPI results essentially the same as the Jacobian.

\subsection*{Supplementary Note 5: Implementation of competing methods}\label{supp-note3}

In our study, the EC inference performance is compared with two competing methods: Granger causality and dynamic causal modeling (DCM). The signal prediction performance is compared with a baseline model: univariate auto-regression of order 3 (i.e., UAR(3)).

\subsubsection*{Granger causality test}

Granger causality is based on the principle that causal relationships in time series data can be discerned through predictability. If the prediction of one time series is enhanced by the inclusion of past values from another, the latter is said to ``Granger-cause" the former. 

Granger causality in neuroimaging is to test whether the past values of brain activity in one region can predict future activity in another, beyond the predictive capacity of past values of the latter region alone. This is operationalized using vector autoregressive (VAR) models, where each component of the vector $\mathbf{y}_t$ represents the activity at a specific brain region at time $t$. The VAR model is formulated as follows,
\begin{equation}
\mathbf{y}_t = \sum_{i=1}^p \mathbf{A}_i \mathbf{y}_{t-i} + \mathbf{e}_t,
\end{equation}
where the matrices $\mathbf{A}_i$ contain the coefficients that describe the influence of the activity in all brain regions at previous time points $t-i$ on the current activity at time $t$, and $p$ is the number of lags. The term $\mathbf{e}_t$ is the error vector, representing unpredicted variations in the signal. The significance of each coefficient in $\mathbf{A}_i$ reveals the directional influence between regions, indicating whether past activity in one or more regions can predict current activity in another. 

We implemented the Granger causality in Python based on the algorithm in MVGC MATLAB toolbox ~\cite{Barnett2014MVGC}. The number of lags $p$ is set to 3.

\subsubsection*{Dynamic causal modelling}

Dynamic causal modeling (DCM) for fMRI is based on a generative model with two components~\cite{Friston2014DCM}. First, a neuronal model describes interactions among neuronal populations. Second, neuronal activities are mapped to observed hemodynamic responses. 

The state equation of the neuronal model is defined as
\begin{equation}
\dot{\mathbf{x}}(t) = \mathbf{A} \mathbf{x}(t) + \eta,
\end{equation}
where $\mathbf{x}(t)$ represents the vector of hidden neuronal states for each region at time $t$. $\mathbf{A}$ is the matrix describing the intrinsic connectivity between regions. $\eta$ is the noise, representing intrinsic neuronal fluctuations.

The hemodynamic response is modeled using a function $h$, transforming the neuronal states into BOLD signals,
\begin{equation}
    \mathbf{y}(t)=h(\mathbf{x}(t))+\epsilon ,
\end{equation}
where $h(\cdot)$ is the nonlinear hemodynamic response function. $\epsilon$ represents measurement noise in the observed BOLD signal.

For resting-state data, DCM is often implemented in the frequency domain to fit the observed cross-spectral density in terms of the parameters in $\mathbf{A}$, which describes the directional influences between brain regions. Bayesian inference is typically employed to fit the model parameters. We implemented DCM using function `\textit{spm\_dcm\_fmri\_csd}' of SPM 12 in MATLAB.

\subsubsection*{Univariate auto-regression}

A univariate auto-regression (AR) model of order $p$ (i.e., AR($p$)) is expressed as follows:
\begin{equation}
y_t=c+a_1y_{t-1}+a_2y_{t-2}+\ldots+a_py_{t-p}+\epsilon_t,
\end{equation}
where $y_t$ is the value of the time series at time $t$, $c$ is a bias constant, $a_1,a_2,\ldots,a_p$ are the parameters of the model which represent the impact of the $p$ previous time points on the current value. $\epsilon_t$ is the error term at time $t$.

The parameters $c$ and $a_1,a_2,\ldots,a_p$ are estimated separately for each brain region. Optimization was performed using gradient descent with Adam optimizer and learning rate $10^{-3}$.  \\

\subsection*{Supplementary Note 6: Applying NPI on ADNI and ABIDE datasets}\label{supp-note4}

For the ABIDE dataset, we used rs-fMRI data with a repetition time (TR) of \re{2 seconds} from 234 autism patients and 285 healthy controls. Data preprocessing was performed using \textit{SPM8} toolbox~\cite{penny2011statistical}. The preprocessing steps included slice timing correction, realignment and reslicing to correct for motion, coregistration of BOLD images to anatomical sequences, normalization to a standard template, segmentation of brain tissues, bandpass filtering (0.001-0.1 Hz), and regression of nuisance variables (e.g., motion parameters and physiological noises) from CSF and white matter~\cite{craddock2013neuro}.

For the ADNI dataset, we randomly selected rs-fMRI data with a TR of 3 seconds from 60 patients with Alzheimer's disease and 60 healthy controls. Data preprocessing was conducted using \textit{DPABI8.2} toolbox~\cite{yan2016dpabi}. The preprocessing steps included removing the first 10 volumes, slice timing correction, field map-based distortion correction, and head motion correction. The individual functional images were then coregistered and spatially normalized to the EPI template. The normalized images were smoothed with a Gaussian kernel with an FWHM of 4 mm. Finally, linear detrending and nuisance signal regression were applied.

In the analysis of ABIDE and ADNI datasets, the brain is parcellated into 39 regions according to the Multi-Subject Dictionary Learning (MSDL) atlas~\cite{Varoquaux2011Multisubject}. This atlas, characterized by probabilistic regions, allows for overlapping among the ROIs, reflecting the complex and intermingled nature of brain functionalities. These ROIs encompass both cortical and subcortical areas, providing a comprehensive framework for investigating effective connectivity and brain dynamics. 

Due to the limited length of fMRI data in these two datasets, we employed a two-stage training strategy to optimize an MLP model for the prediction of fMRI data. In stage 1, we pre-train the MLP using the fMRI data from all subjects. In stage 2, we fine-tune the MLP on individual subject data. During pre-training, all model parameters are trainable, allowing the MLP to learn feature extraction and make effective predictions from fMRI data, thereby capturing the general patterns and structures within the data. In the fine-tuning stage, only the parameters in the final layer of the MLP are trainable. This enables the model to adjust its output to more accurately reflect the fMRI signals of individual subjects while preserving the feature extraction mechanisms learned during pre-training. Afterward, we apply perturbations to each individualized ANN model to derive the subject-specific EC.

\newpage
\renewcommand{\figurename}{Supplementary Fig.}
\subsection*{Supplementary figures}
\setcounter{figure}{0} 

\noindent
Supplementary Fig. 1: Optional surrogate model architectures.  \\
Supplementary Fig. 2: Signal changes after perturbing a node of RNN. \\
Supplementary Fig. 3: Ground-truth EC and NPI-inferred EC of RNN. \\
Supplementary Fig. 4: NPI outperforms Granger causality (GC) in EC inference from ground-truth RNN generated data. \\
Supplementary Fig. 5: EC from NPI and GC on the synthetic data generated by WBM. \\
Supplementary Fig. 6: NPI-inferred EC on the left hemisphere using WBM-generated BOLD data. \\
Supplementary Fig. 7: Comparison across different prediction models on WBM-generated data. \\
Supplementary Fig. 8: Performance comparison between single-step-input and three-step-input ANN. \\
Supplementary Fig. 9: NPI-inferred EBC from the HCP dataset using the AAL atlas. \\
Supplementary Fig. 10: The EC with the strongest strengths and regions with the largest degrees. \\
Supplementary Fig. 11: $p$-values of EC inferred from the HCP dataset. \\
Supplementary Fig. 12: Applying NPI to the ABIDE dataset and ADNI dataset. \\
Supplementary Fig. 13: EC obtained by incorporating hemodynamic convolution in ANN perturbation. \\
Supplementary Fig. 14: EC obtained by positive/negative impulse perturbation to the surrogate model. \\
Supplementary Fig. 15: Correlation of intra-network FC of seven functional networks across sessions, subjects, and datasets. \\

\newpage
\begin{figure}[H]
\label{fig:CNN}
\centering
\includegraphics[width=0.9\linewidth]{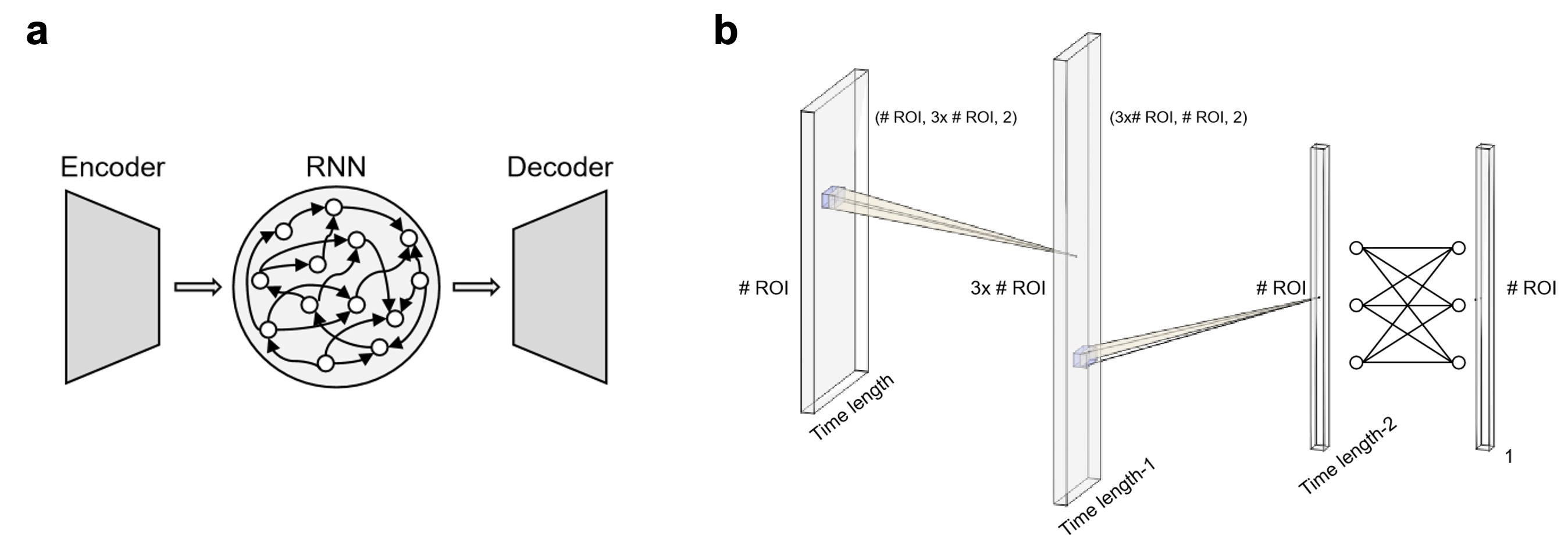}
\caption{\re{\textbf{Optional surrogate model architectures.}
\textbf{a}, Architecture of the recurrent neural network (RNN). A generic state-space model contains an encoder module, a recurrent neural network, and a decoder module.
\textbf{b}, Architecture of the convolutional neural network (CNN). The designed CNN contains a two-layer one-dimensional (1D) convolution, with a Relu activation function between them. A linear layer maps the output of the last convolution layer to the predicted time series. The parameters of 1D convolution are represented as (input channels, output channels, kernel size).}
}
\label{figS:RNN-CNN}
\end{figure}

\newpage
\begin{figure}[H]
\centering
\includegraphics[width=0.9\linewidth]{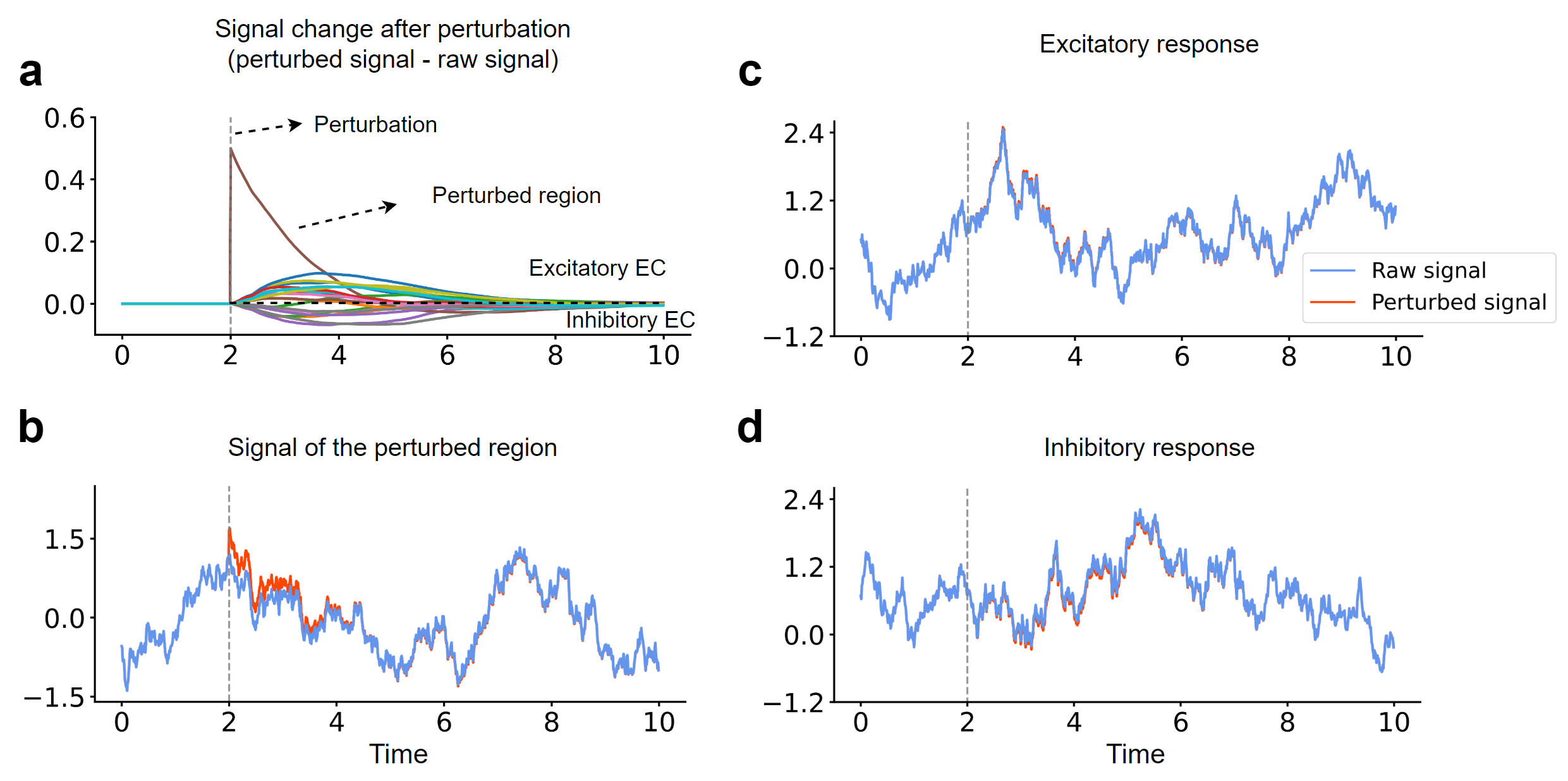}
\caption{\textbf{Signal changes after perturbing a node of RNN.}   
\textbf{a}, The perturbation-induced signal changes are calculated as the difference between the generated signal with and without perturbation. Regions that have excitatory EC from the perturbed region exhibit positive responses. Regions that have inhibitory EC from the perturbed region induce negative responses.
\textbf{b}, Signal of the perturbed region.
\textbf{c}, Signal of a region that has excitatory EC from the perturbed region.
\textbf{d}, Signal of a region that has inhibitory EC from the perturbed region.
}
\label{figS:pert}
\end{figure}

\newpage
\begin{figure}[H]
\centering
\includegraphics[width=0.9\linewidth]{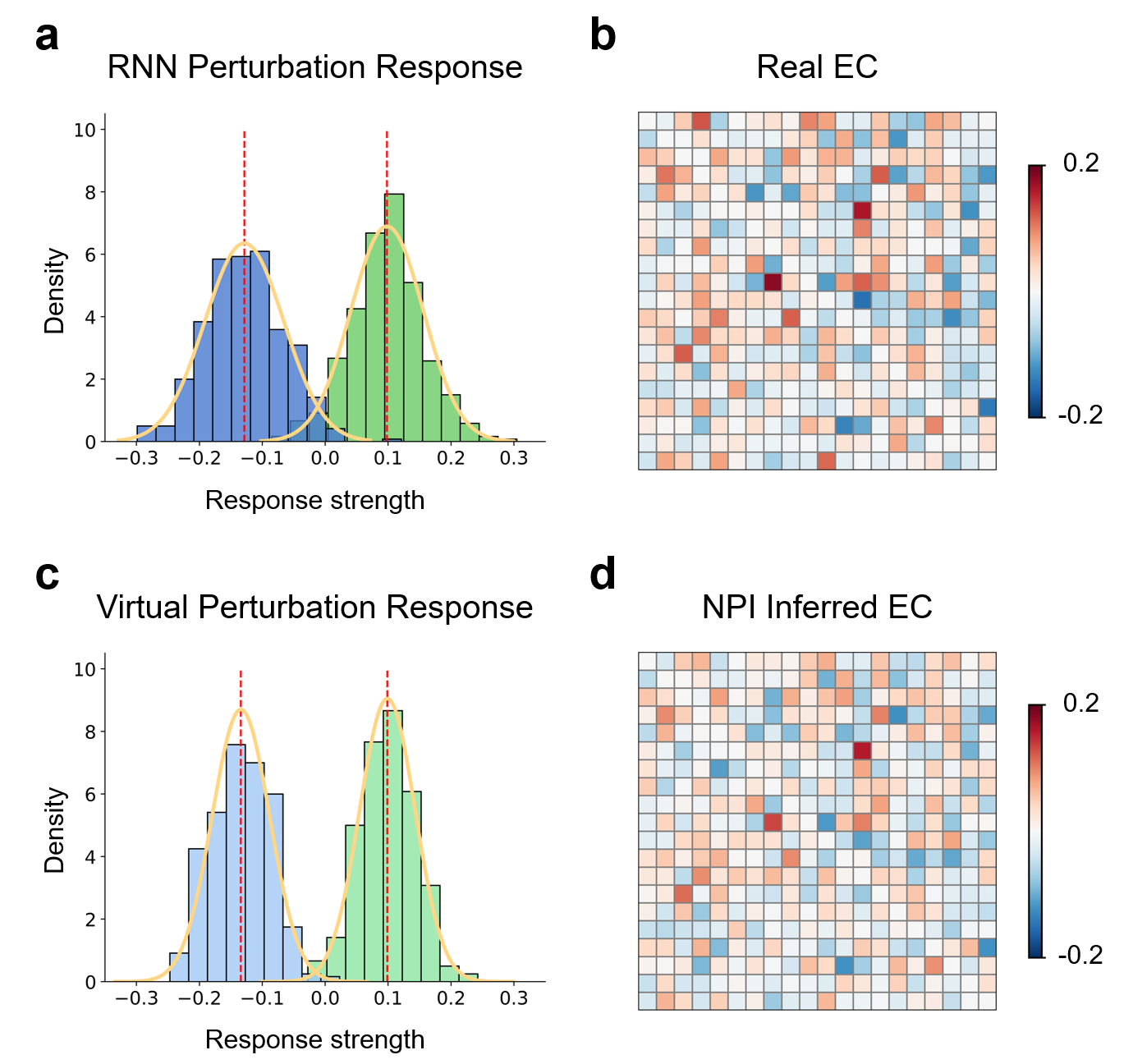}
\caption{\textbf{Ground-truth EC and NPI-inferred EC of ground-truth RNN.} 
\textbf{a}, The distribution of response of node 11 after perturbing node 17 in RNN is shown in blue. The mean response magnitude is -0.13 and STD is 0.06. The response of node 1 after perturbing node 3 in RNN is shown in green. The mean response magnitude is 0.10 and STD is 0.06. 
\textbf{b}, The real EC is calculated as the mean response after perturbation. 
\textbf{c}, The distribution of response after perturbing the same nodes as (a) but in ANN. The blue one has a mean value of -0.13 and an STD of 0.05. The green one has a mean value of 0.10 and std 0.04. 
\textbf{d}, EC inferred by NPI as the mean response after perturbing ANN.
}
\end{figure}

\newpage
\begin{figure}[H]
\centering
\includegraphics[width=0.9\linewidth]{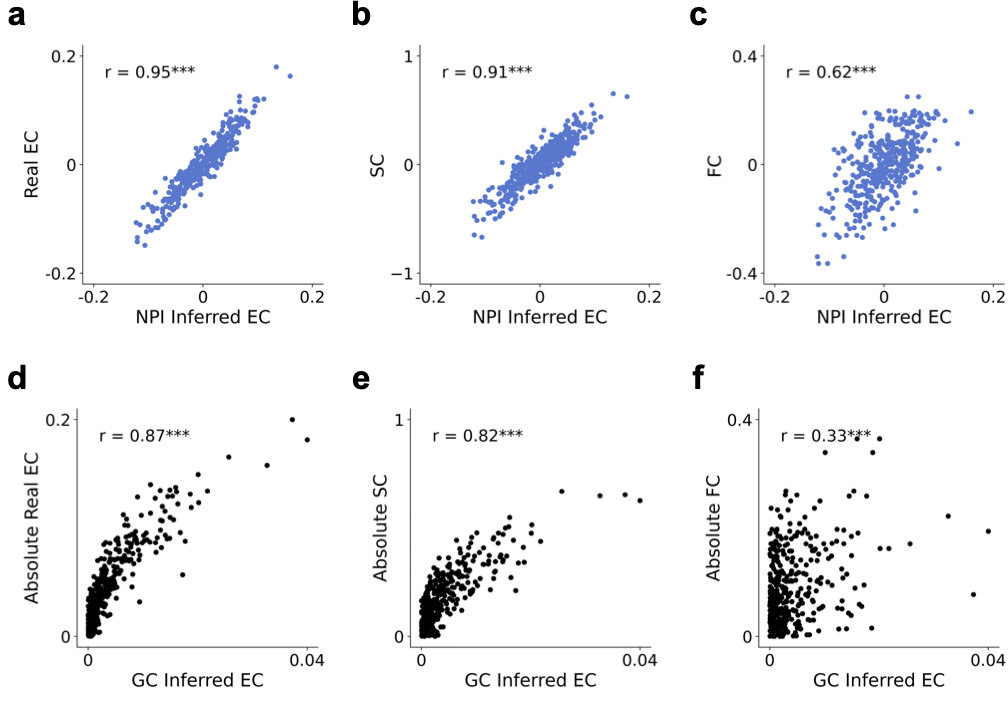}
\caption{\textbf{NPI outperforms Granger \re{causality} (GC) in EC inference from ground-truth RNN generated data.}  
\textbf{a-c}, The relationship between NPI-inferred EC and ground-truth EC (a), SC (b), and FC (c) of RNN.
\textbf{d-f}, The relationship between GC-inferred EC and ground-truth EC (d), SC (e), and FC (f) of RNN. The results show that NPI-inferred EC has higher correlations with ground-truth EC, SC, and FC, compared to those from GC-inferred EC.
}
\end{figure}

\newpage
\begin{figure}[H]
\centering
\includegraphics[width=0.9\linewidth]{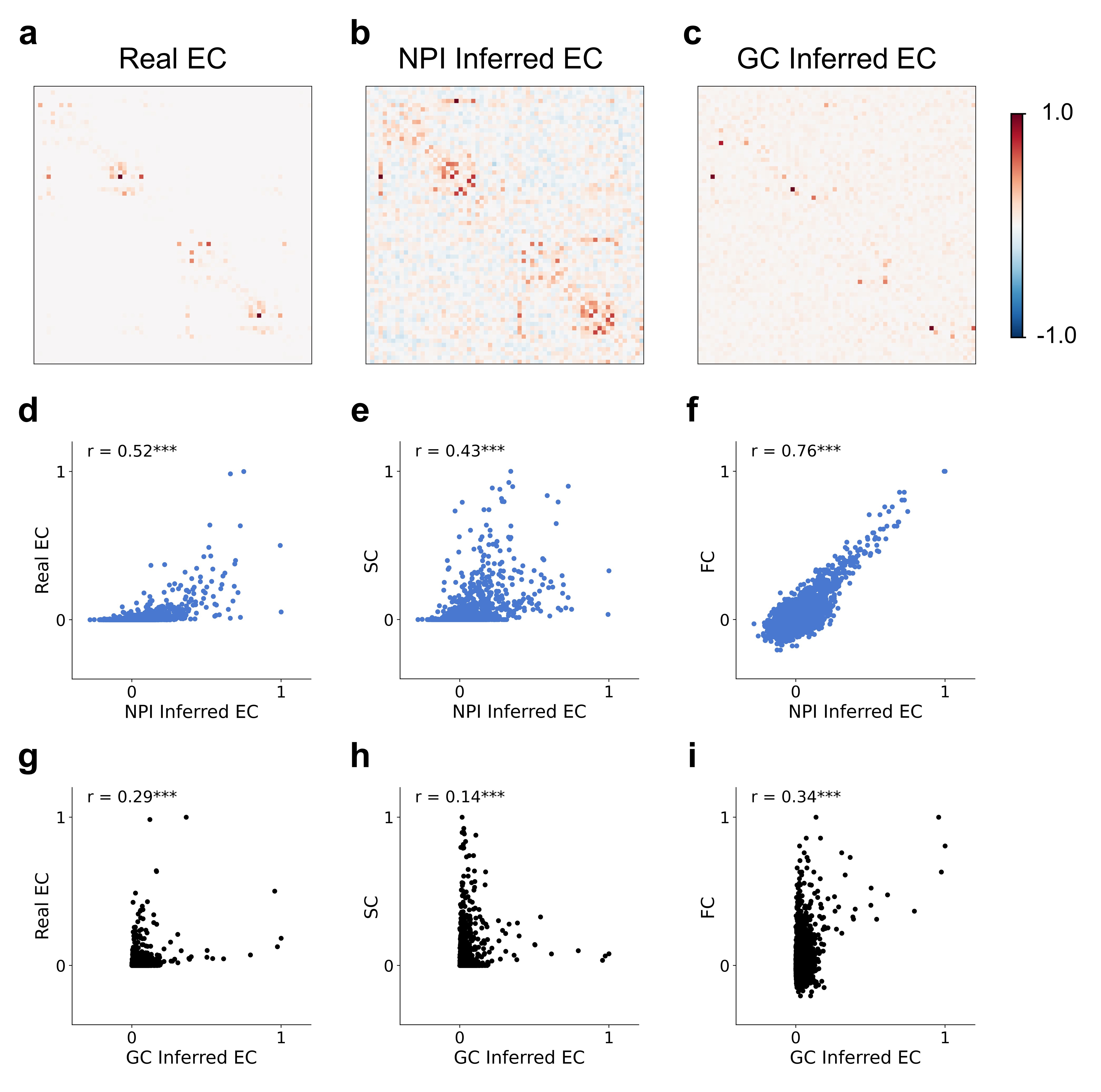}
\caption{\textbf{EC from NPI and GC on the synthetic data generated by WBM.}  
\textbf{a}, Ground-truth EC obtained as the stimulation-induced response in WBM.
\textbf{b}, EC inferred by NPI.
\textbf{c}, EC inferred by GC.
\textbf{d-f}, The relationship between NPI-inferred EC and ground-truth EC (a), SC (b), and FC (c) of WBM.
\textbf{g-i}, The relationship between GC-inferred EC and ground-truth EC (a), SC (b), and FC (c) of WBM.
}
\end{figure}

\newpage
\begin{figure}[H]
\centering
\includegraphics[width=0.9\linewidth]{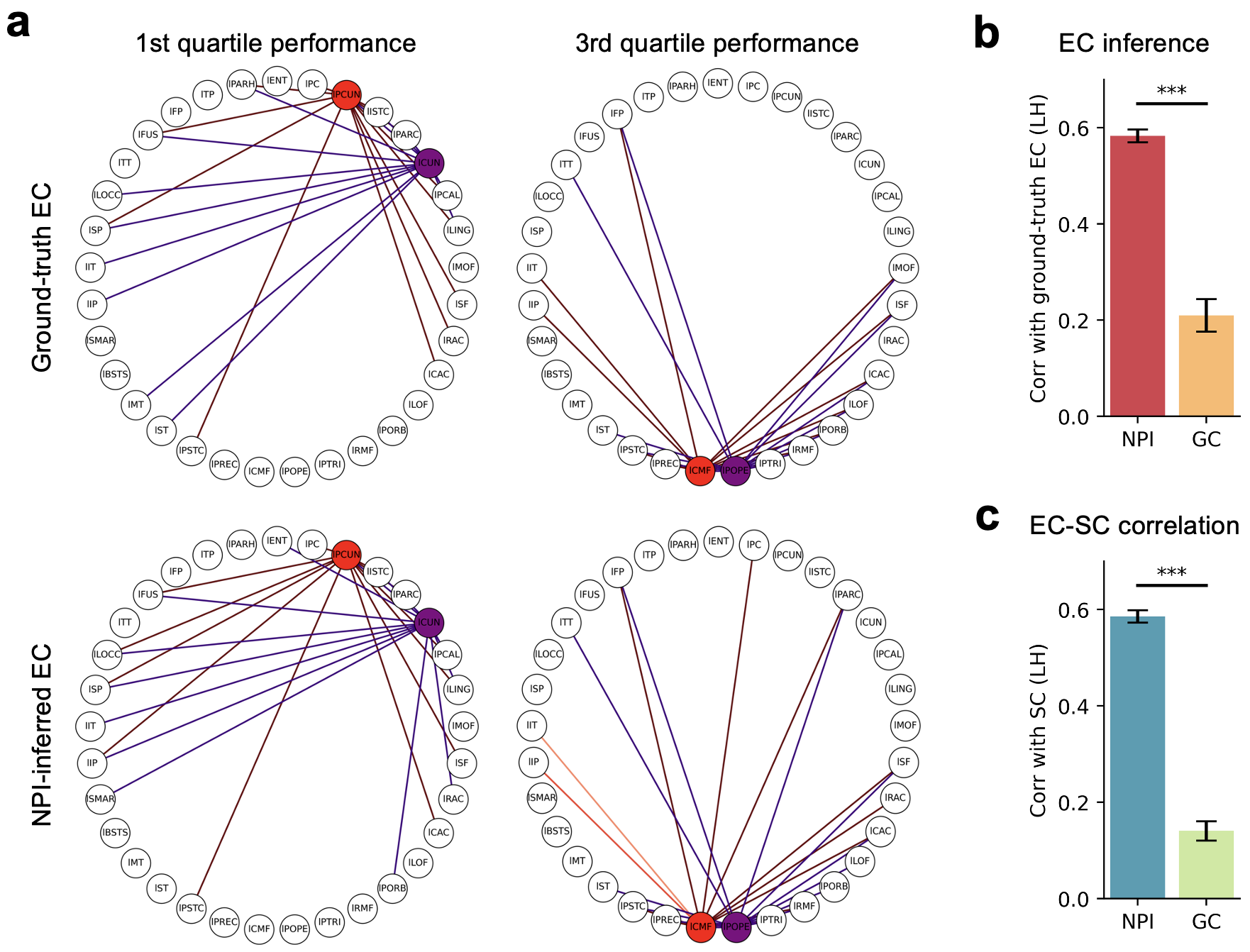}
\caption{\re{\textbf{NPI-inferred EC on the left hemisphere using WBM-generated BOLD data.} 
\textbf{a}, Comparison between ground-truth EC and NPI-inferred EC. Output EC from nodes with first and third-quartile performance are plotted. 
\textbf{b}, \textbf{c}, NPI outperforms GC in capturing the characteristics of left-hemisphere (LH) EC (o) and SC (p) ($p < 10^{-3}$, Wilcoxon signed-rank test). Error bars represent standard deviation.
}}
\end{figure}

\newpage
\begin{figure}[H]
\centering
\includegraphics[width=0.6\linewidth]{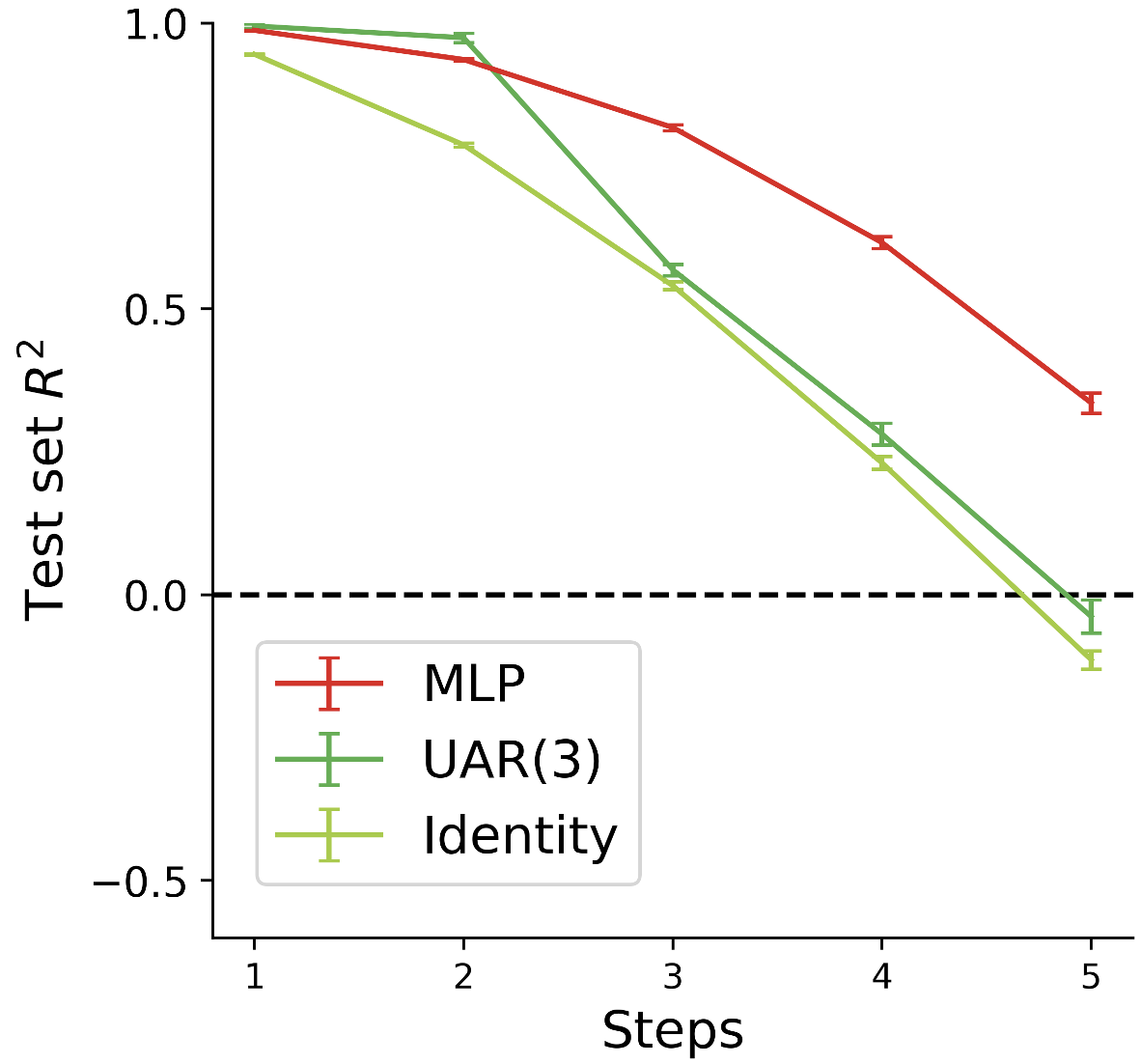}
\caption{\re{\textbf{Comparison across different prediction models on WBM generated data.}
\textbf{a}, Signal prediction performance on the test set of WBM-generated data. Data from 100 WBMs is used. Each data has a length of 10000 TR (TR = 0.72 s). Three models are tested: MLP, UAR, and an identity-output model (predict $x(t+\Delta t)$ to be the same as $x(t)$).
}}
\end{figure}

\begin{figure}[H]
\centering
\includegraphics[width=1\linewidth]{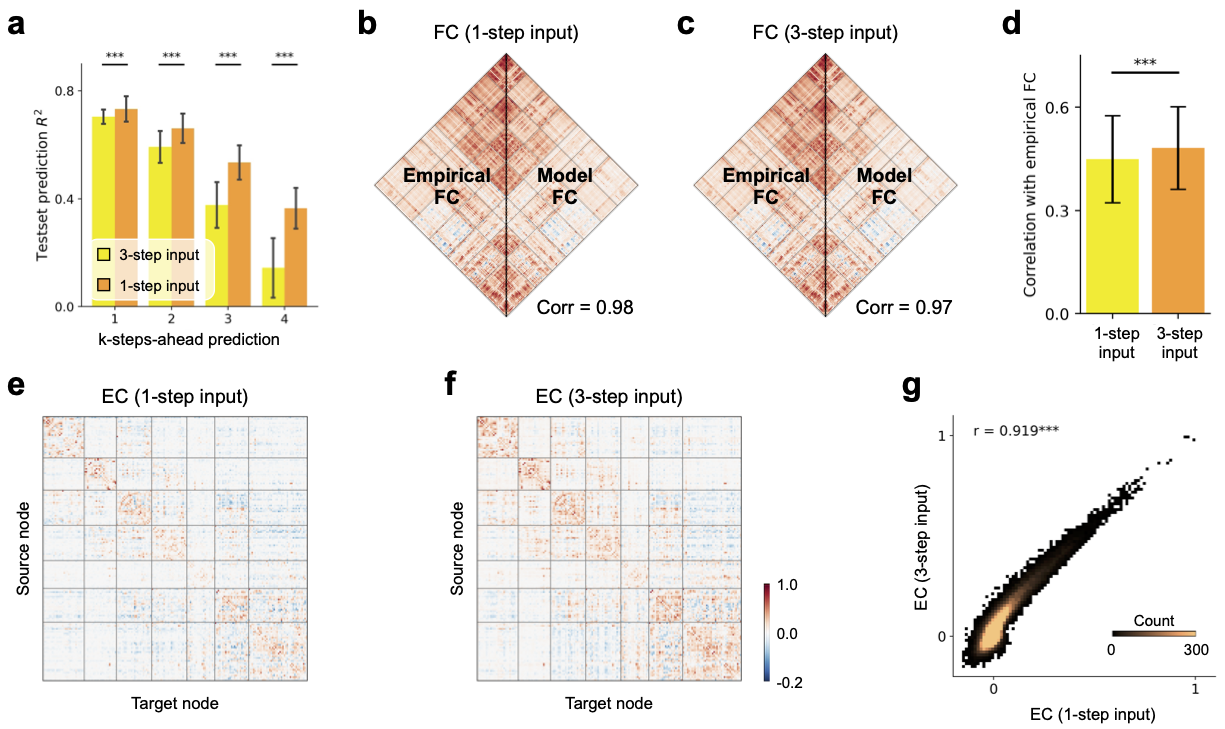}
\caption{\re{\textbf{Performance comparison between single-step-input and three-step-input ANN}. 
\textbf{a}, Test set prediction performance ($R^2$) across individual data from the HCP dataset using the MMP atlas with 360 regions. 
We compare the prediction performance of MLP by using the preceding single step and the preceding three steps as input to predict the next step.
The three-step input consistently outperforms the single-step input in predictive accuracy.
\textbf{b, c}, Comparison of group-level empirical FC with FC reproduced by ANNs trained using preceding one step (b) and three steps (c). Both settings can accurately reproduce group-level FC, averaged across 800 subjects.
\textbf{d} Correlation between individual-level empirical FC and FC reproduced by trained ANN, highlighting that the three-step input model demonstrates significantly higher correlation than the single-step model.
\textbf{e,f} EC derived by perturbing ANNs trained with single-step (e) and three-step (f) inputs.
\textbf{g} Correlation of EC derived from single-step-input and three-step-input ANNs, showing a high correlation ($r=0.919, p < 10^{-3}$) between the two approaches, indicating robustness across different training methods. The color bar (0-300) indicates the density of EC pairs.
}}
\label{fig:1step}
\end{figure}

\newpage
\begin{figure}[H]
\centering 
\includegraphics[width=1\linewidth]{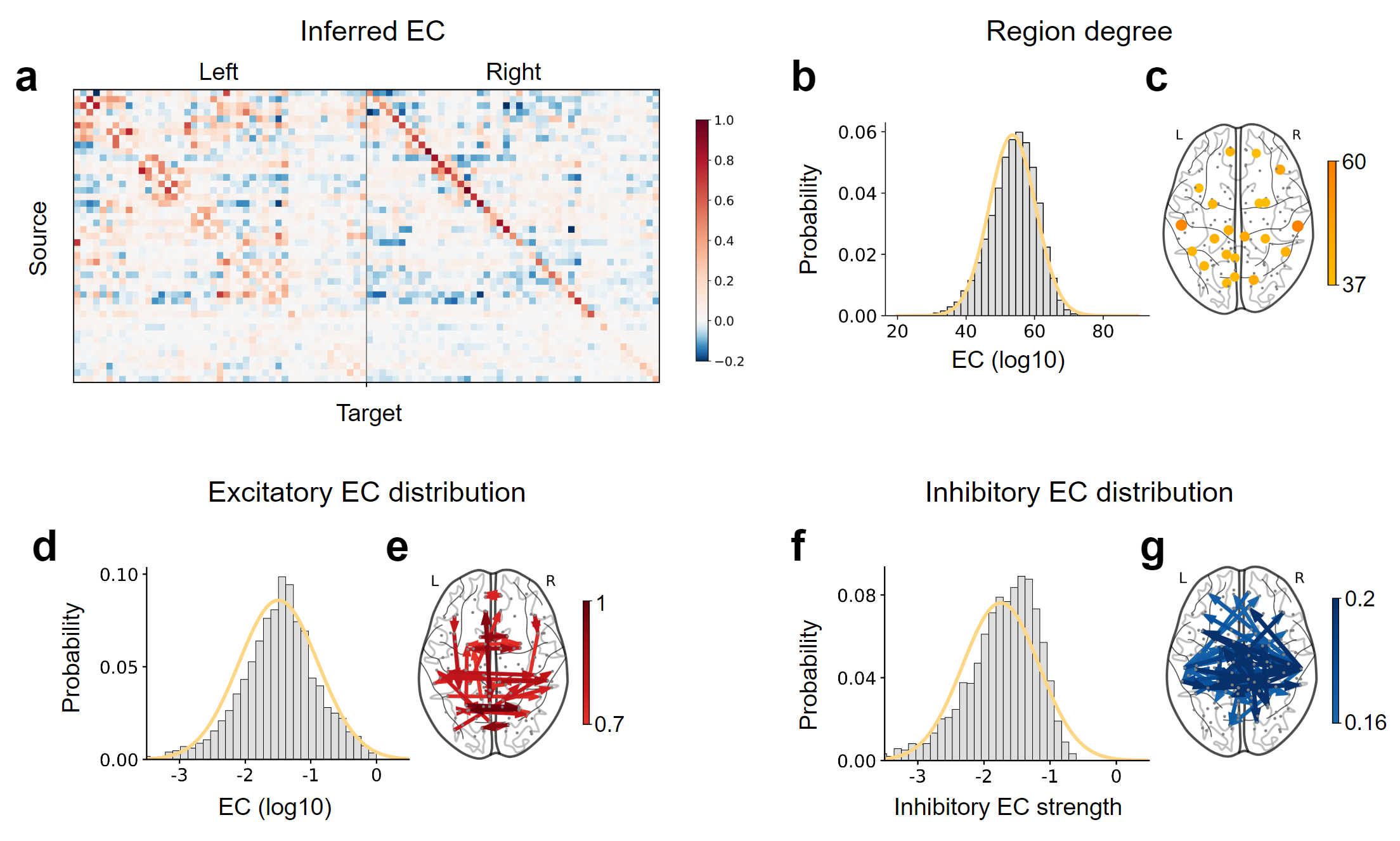}
\caption{\textbf{NPI-inferred EBC from the HCP dataset using the AAL atlas.} 
(\textbf{a}) The inferred EC from the left hemisphere to the whole brain. 
(\textbf{b}) The degree distribution of regions in the binarized EBC. EBC is binarized by a threshold larger than 80\% of EC. The degree is calculated as the mean of the in-degree and out-degree of each region. 
(\textbf{c}) 20 brain regions with the largest degree after binarizing EBC.
(\textbf{d, f}) The distribution of excitatory (d) and inhibitory (f) EC across the whole brain. The distribution of both can be the best fit by a log-normal distribution. The fitting curves of Gaussian distribution to the log-transformed EC strengths are plotted in yellow. 
(\textbf{e,g}) The 50 strongest excitatory (e) and inhibitory (g) EC. 
}
\end{figure}

\newpage
\begin{figure}[H]
\centering
\includegraphics[width=1\linewidth]{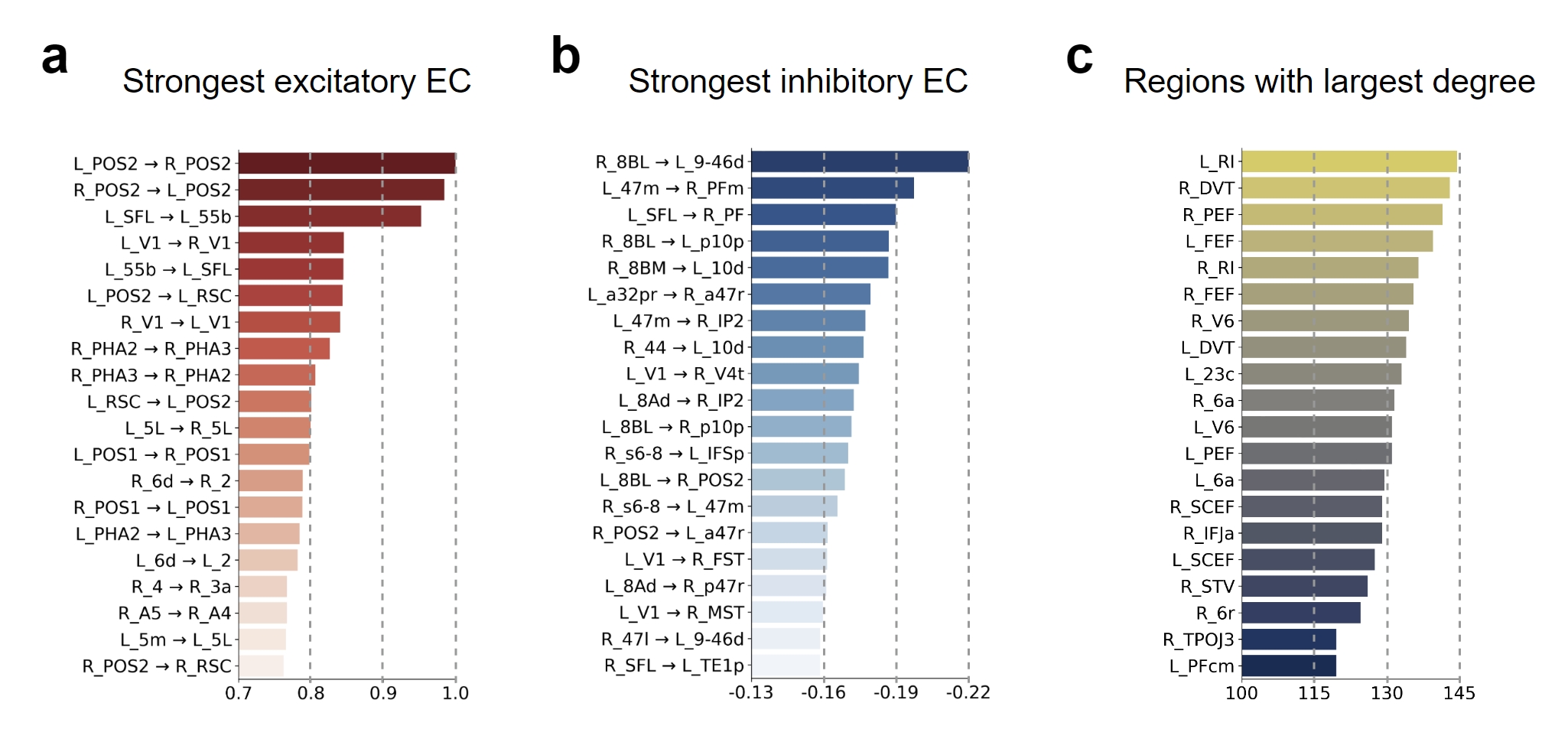}
\caption{\textbf{The EC with strongest strengths and regions with largest degrees.} (\textbf{a,b}) The top 20 strongest excitatory (a) EC and inhibitory (b) EC. (\textbf{c}) 20 brain regions with the largest degree after binarizing EBC.}
\label{fig:strongestec}
\end{figure}

\newpage
\begin{figure}[H]
\centering
\includegraphics[width=0.6\linewidth]{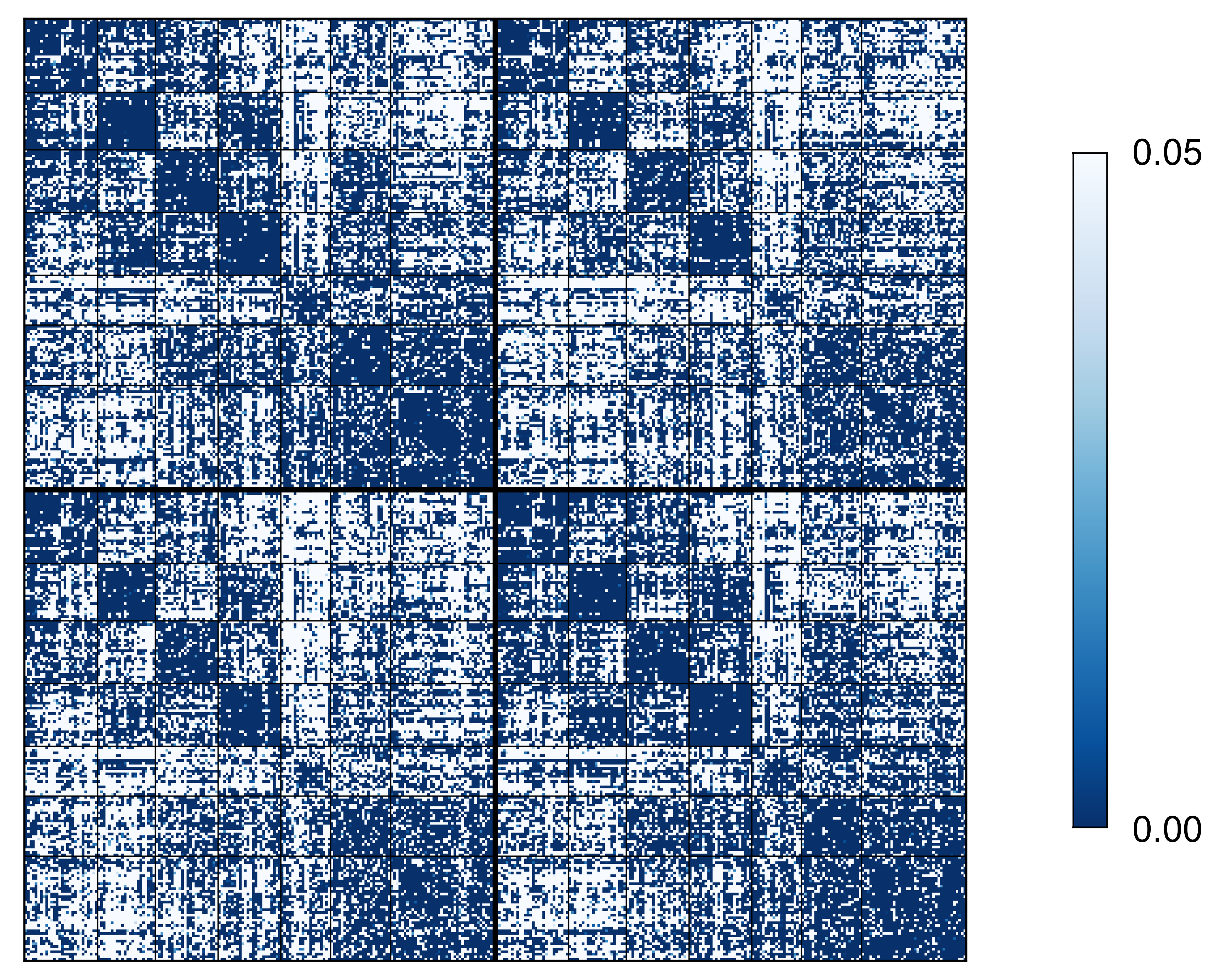}
\caption{\textbf{$p$-values of EC inferred from HCP dataset.} Analysis of EC across 800 subjects, employing a single-sample $t$-test with Bonferroni correction, reveals that 55\% of EC entries significantly differ from zero.
}
\label{fig:pvalue}
\end{figure}

\newpage
\begin{figure}[H]
\centering
\includegraphics[width=1\linewidth]{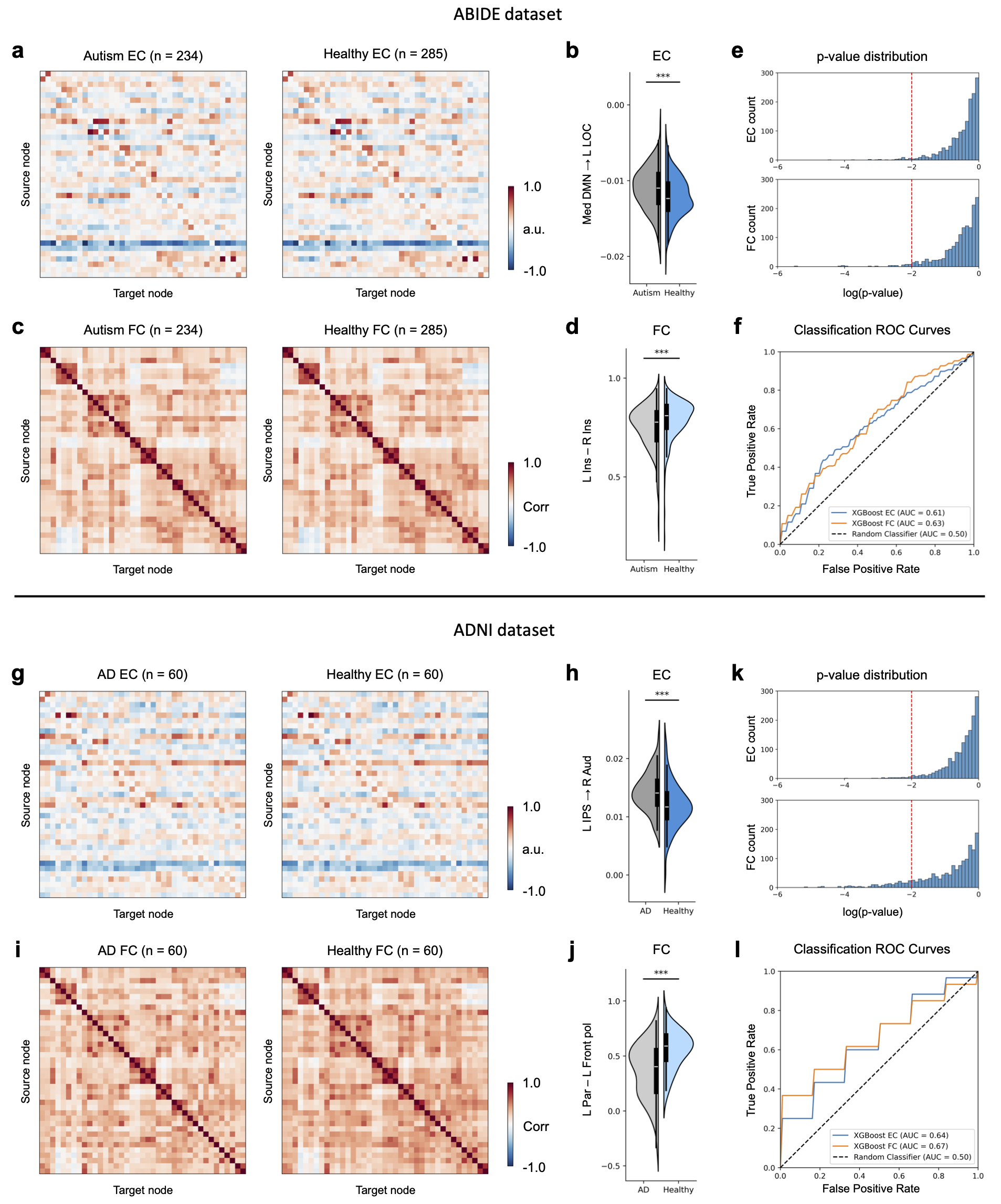}
\caption{\re{\textbf{Applying NPI to ABIDE dataset and ADNI dataset}. 
\textbf{a-f}, Applying NPI to rs-fMRI data from patients with autism disease. The rs-fMRI data from 234 autism patients and 285 healthy controls are randomly selected from the autism brain imaging data exchange (ABIDE) dataset. MSDL atlas with 39 regions was used for parcellation.
\textbf{a}, NPI-inferred EC from rs-fMRI data of autism patients (left) and healthy controls (right).
\textbf{b}, The EC connection that is most significantly different between autism patients and healthy controls. 
\textbf{c}, FC calculated from rs-fMRI data of autism patients (left) and healthy controls (right).
\textbf{d}, The FC connection that is most significantly different between autism patients and healthy controls.
\textbf{e}, Distribution of log(p-value) of EC connection differences between autism patients and healthy subjects.
\textbf{f}, Classification ROC curve of using EC or FC to classify autism patients and healthy controls.
\textbf{g-i}, the same as a-f, for the Alzheimer’s Disease Neuroimaging Initiative (ADNI) dataset. The rs-fMRI data from 60 AD and 60 healthy controls are randomly selected.
}}
\label{fig:adni}
\end{figure}

\newpage
\begin{figure}[H]
\centering
\includegraphics[width=0.9\linewidth]{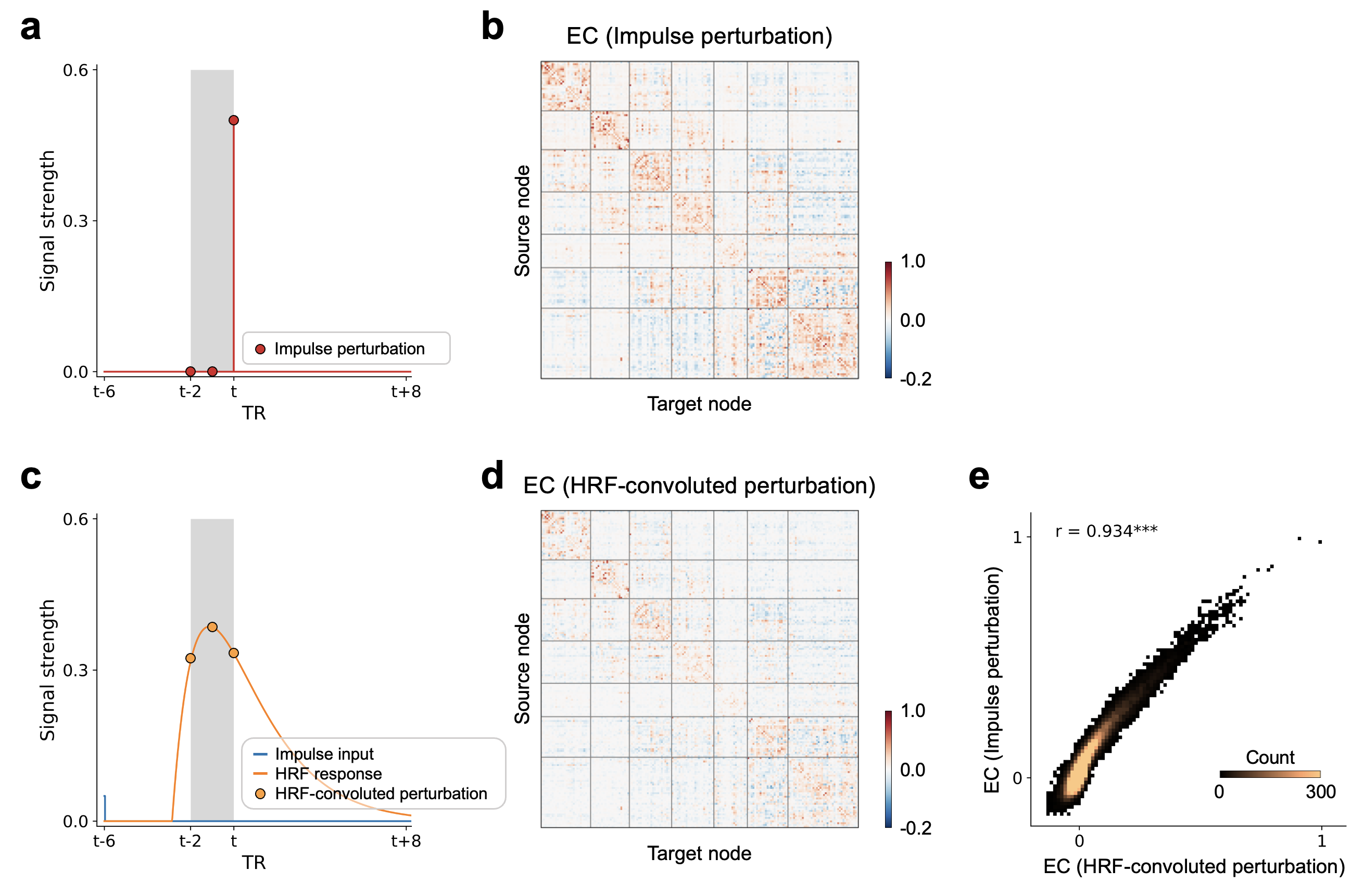}
\caption{\re{\textbf{EC obtained by incorporating hemodynamic convolution in ANN perturbation.} 
\textbf{a}, In regular NPI, an impulse perturbation is applied at only time $t$. The response at time $t+1$ is calculated as EC.
\textbf{b}, Group-level EC from HCP dataset with MMP atlas inferred by applying an impulse perturbation.
\textbf{c}, Perturbation signal incorporating hemodynamic response. An impulse of short duration (0.02 s) and strength 0.05 was applied at t-6 TR. The resulting hemodynamic response function (HRF) is modeled as the convolution of this impulse with a hemodynamic kernel (Supplementary Note 3). The responses at time t-2, t-1, and t TR are used as inputs for perturbation in the ANN.
\textbf{d}, Group-level EC from the HCP dataset with MMP atlas inferred using the HRF-convoluted impulse as the perturbation to the trained ANN.
\textbf{e}, Correlation between EC derived from the original impulse perturbation and the HRF-convoluted perturbation, with a correlation coefficient of $r=0.934$, indicating a strong alignment between the two methods.}}
\label{fig:hrf}
\end{figure}

\begin{figure}[H]
\centering
\includegraphics[width=0.9\linewidth]{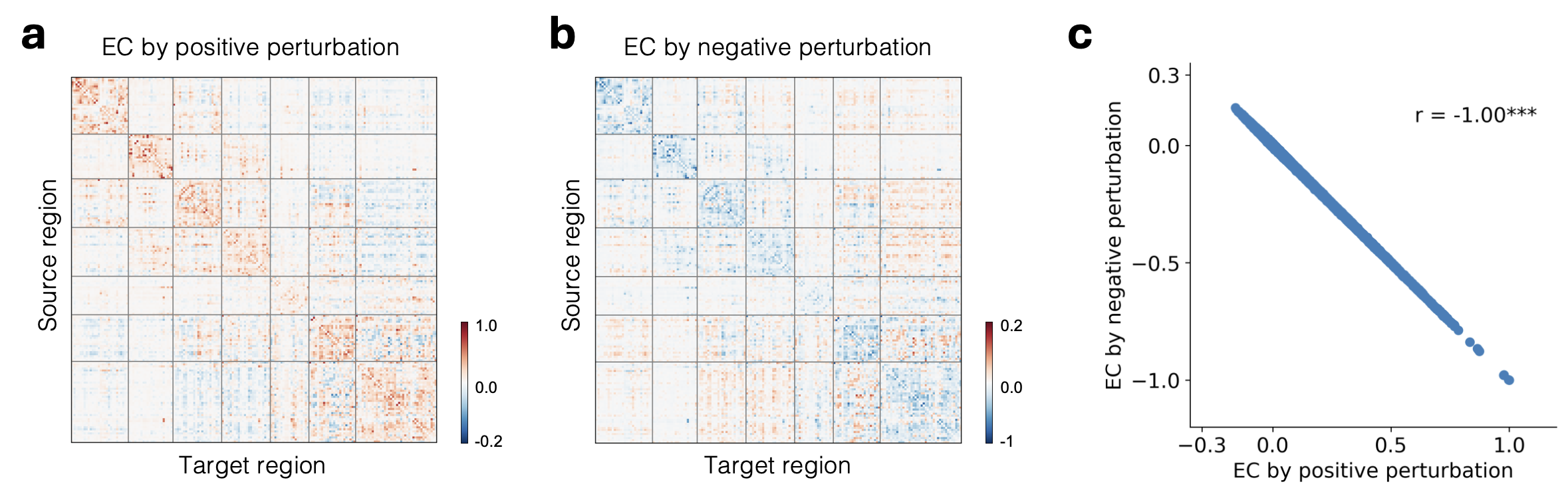}
\re{\caption{\textbf{EC obtained by positive/negative impulse perturbation to the surrogate model.} 
\textbf{a}, NPI-inferred EC of the left hemisphere. It is obtained by applying a positive perturbation.
\textbf{b}, Response of negative perturbation. Negative perturbations with the same magnitude as in regular positive perturbations but reversed signs were applied.
\textbf{c}, Correlation analysis between the ANN's response to the positive and negative perturbation ($r=-1.00$, $p<1e-3$).}}
\label{R1-negpert}
\end{figure}

\newpage
\begin{figure}[H]
\centering
\includegraphics[width=0.7\linewidth]{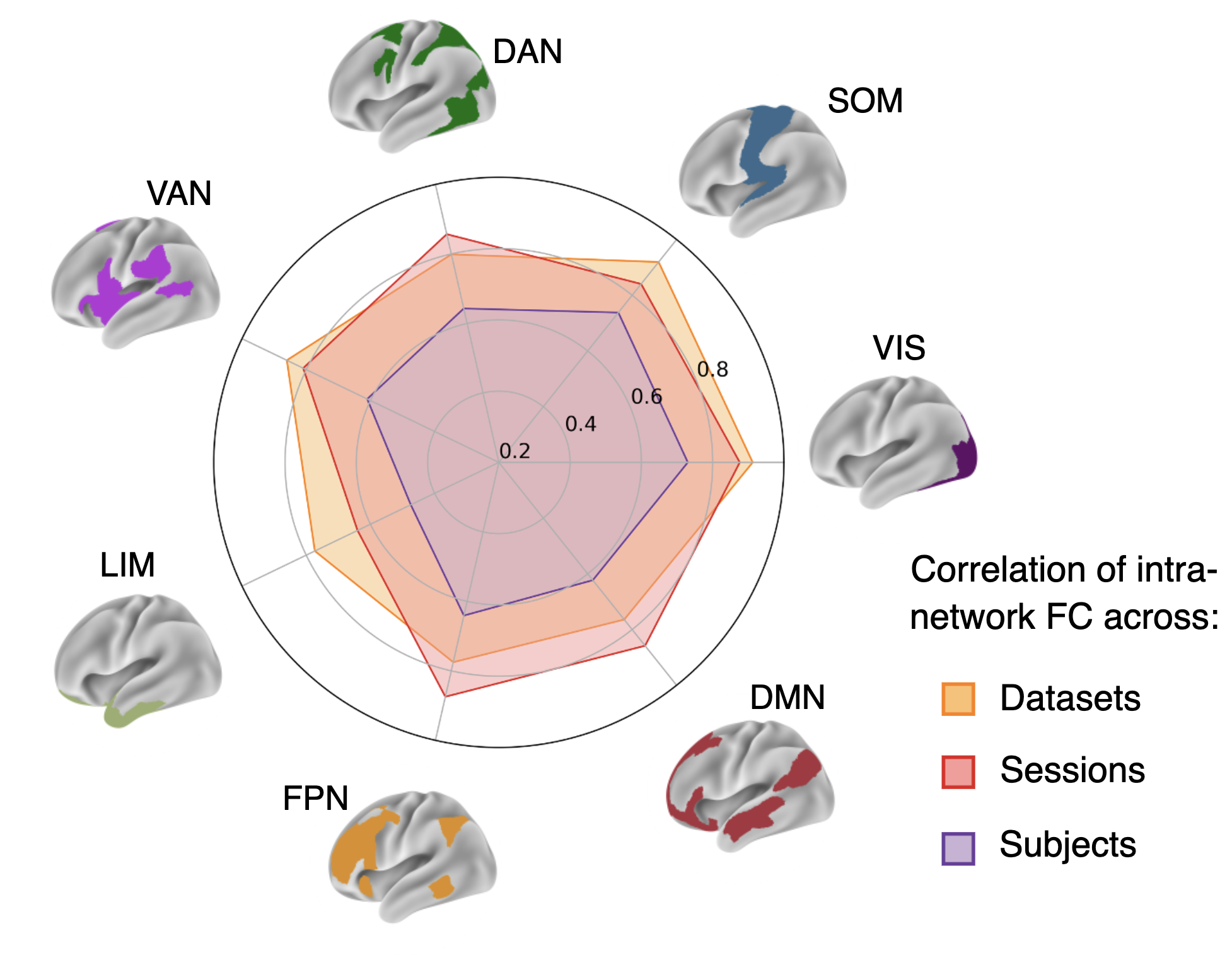}
\caption{\textbf{Correlation of intra-network FC of seven functional networks across sessions, subjects, and datasets.} 
Sessions refers to splitting and calculating FC on each half of the individual data with results averaged across 800 subjects. Subjects refers to cross-subject FC correlation from 800 subjects. Datasets refers to the consistency between FC calculated from HCP and ABCD datasets.
}
\label{fig:fc_robustness}
\end{figure}

\newpage
\subsection*{Supplementary tables}
\renewcommand{\tablename}{Supplementary Table}

\noindent
Supplementary Table 1: Performance of one-step-ahead prediction, FC reproduction, and EC inference across surrogate models and datasets. \\
Supplementary Table 2: Correlation between the NPI-inferred EC and the SC across surrogate models and datasets. \\
Supplementary Table 3: MNI Coordinates of seed regions in seven brain functional networks. \\
Supplementary Table 4: Fitting the distribution of EBC strengths to various density functions. \\
Supplementary Table 5: Name and order of the MMP atlas for the left hemisphere. Regions in the right hemisphere are ordered using the same order.

\begin{table}[H]
\caption{Performance of one-step-ahead prediction, FC reproduction, and EC inference across surrogate models and datasets. $\checkmark$ means that the model has the ability to do that but we do not have the ground-truth EC to quantify the EC inference.}
\resizebox{\textwidth}{!}{%
\begin{tabular}{l|l|c|c|c|c}
\hline
\multicolumn{1}{c|}{}                        & \multicolumn{1}{c|}{}         & NPI-MLP                      & NPI-CNN                      & NPI-RNN                & NPI-VAR(3)                   \\ \hline
\multirow{3}{*}{RNN synthetic data}          & Signal Prediction             & 0.25405$\pm$0.02556          & 0.24236$\pm$0.02628          & 0.26412$\pm$0.02525    & \textbf{0.27488$\pm$0.02490} \\
                                             & FC reproduction               & 0.78635$\pm$0.05797          & 0.87575$\pm$0.04285          & 0.82595$\pm$0.04983    & \textbf{0.97511$\pm$0.00587} \\
                                             & EC inference                  & 0.95107$\pm$0.00600          & 0.95573$\pm$0.00581          & 0.94821$\pm$0.00654    & \textbf{0.96523$\pm$0.00462} \\ \hline
\multirow{3}{*}{Open dataset synthetic fMRI} & Signal Prediction             & 0.58668$\pm$0.04629          & 0.59383$\pm$0.04640          & 0.55537$\pm$0.03800    & \textbf{0.59400$\pm$0.04618} \\
                                             & FC reproduction               & 0.71357$\pm$0.15480          & 0.77684$\pm$0.13414          & 0.63137$\pm$0.09915    & \textbf{0.98332$\pm$0.01032} \\
                                             & EC inference                  & $\checkmark$                 & $\checkmark$                 & $\checkmark$           & $\checkmark$                 \\ \hline
\multirow{3}{*}{WBM synthetic fMRI}          & Signal Prediction             & 0.98718$\pm$0.00066          & 0.98887$\pm$0.00048          & 0.89804$\pm$0.00169    & \textbf{0.99319$\pm$0.00011} \\
                                             & FC reproduction               & \textbf{0.54565$\pm$0.02279} & 0.43193$\pm$0.02507          & 0.26836$\pm$0.03516    & 0.08073$\pm$0.02772          \\
                                             & EC inference                  & \textbf{0.53188$\pm$0.00751} & 0.49060$\pm$0.01214          & 0.02091$\pm$0.01870    & 0.52119$\pm$0.00884          \\ \hline
\multirow{3}{*}{Real fMRI}                   & Signal Prediction             & 0.77641$\pm$0.04784          & 0.77476$\pm$0.05094          & 0.78530$\pm$0.04791    & \textbf{0.89711$\pm$0.02985} \\
                                             & FC reproduction               & 0.49275$\pm$0.16381          & \textbf{0.51437$\pm$0.10898} & 0.41622$\pm$0.14159    & 0.20406$\pm$0.09568          \\
                                             & EC inference                  & $\checkmark$                 & $\checkmark$                 & $\checkmark$           & $\checkmark$                 \\ \hline
\end{tabular}}
\end{table}

\begin{table}[H]
\centering
\tiny
\caption{Correlation between the NPI-inferred EC and the SC across surrogate models and datasets.}
\begin{tabular}{c|c|c|cl}
\hline
                  & RNN synthetic   & Open dataset    & WBM synthetic   &  \\ 
\hline                  
Granger causality & 0.87554$\pm$0.01266 & 0.65159$\pm$0.06305 & 0.13345$\pm$0.01047 &  \\
NPI-VAR(3)        & \textbf{0.93064$\pm$0.00533} & 0.65998$\pm$0.07045 & 0.50071$\pm$0.01048 &  \\
NPI-MLP           & 0.91570$\pm$0.00647 & 0.69423$\pm$0.06474 & \textbf{0.53347$\pm$0.00938} &  \\
NPI-CNN           & 0.92005$\pm$0.00558 & \textbf{0.69944$\pm$0.06482} & 0.44471$\pm$0.01338 &  \\
NPI-RNN           & 0.91523$\pm$0.00663 & 0.56341$\pm$0.26790 & 0.01460$\pm$0.01307 & \\
\hline
\end{tabular}
\end{table}

\begin{table}[H]
\scriptsize
\centering 
\caption{MNI Coordinates of seed regions in seven resting-state networks.}
\begin{tabular}{c|p{3cm}|c} 
\hline
Network & Seed region & MNI Coordinate\\
\hline
VIS &  L V2     & (-12.72,  -81.95,   1.04) \\ 
SOM &  L 3b     & (-41.40,  -21.76,  50.97) \\ 
VAN &  L AIP    & (-37.76,  -44.82,  42.39) \\ 
DAN &  L FOP3   & (-37.44,    2.97,  12.33) \\ 
LIM &  L TGv    & (-41.18,   -2.96, -44.84) \\
FPN &  L a9-46v & (-40.90,   49.90,  8.64)  \\ 
DMN &  L 31pv   & (-9.41,   -45.69,  33.04) \\ 
\hline
\end{tabular}
\end{table}

\begin{table}[H]
\scriptsize
\centering 
\caption{Fitting the distribution of EBC strengths to various density functions. A lower AIC value indicates a model that balances goodness of fit with simplicity. Result shows that the EBC follows a log-normal distribution.}
\begin{tabular}{c|c|c|c} 
\hline
Distribution & Excitatory EC & Inhibitory EC & Absolute EC\\
\hline
Normal & 11256 & 1621 & 14598\\ 
Exponential &  2111 & 164 & 2592\\
Inverse-Gaussian &  956 & 41 & 1076\\ 
Log-normal &  \textbf{773} & \textbf{4} & \textbf{832}\\ 
\hline
\end{tabular}
\end{table}

\begin{table}[H]
\centering
\caption{Name and order of the MMP atlas for the left hemisphere. Regions in the right hemisphere are ordered using the same order.}
\begin{tabular}{l|l|l|l|l|l|l|l}
\hline
1  & L\_V1    & 46 & L\_A5    & 91  & L\_FOP3   & 136 & L\_IP2   \\
2  & L\_V6    & 47 & L\_Ig    & 92  & L\_FOP2   & 137 & L\_IP1   \\
3  & L\_V2    & 48 & L\_MBelt & 93  & L\_TPOJ1  & 138 & L\_PFm   \\
4  & L\_V3    & 49 & L\_LBelt & 94  & L\_TPOJ2  & 139 & L\_p47r  \\
5  & L\_V4    & 50 & L\_A4    & 95  & L\_PFop   & 140 & L\_a32pr \\
6  & L\_V8    & 51 & L\_MST   & 96  & L\_PF     & 141 & L\_SFL   \\
7  & L\_V3A   & 52 & L\_FEF   & 97  & L\_PoI1   & 142 & L\_PCV   \\
8  & L\_V7    & 53 & L\_PEF   & 98  & L\_FOP5   & 143 & L\_STV   \\
9  & L\_FFC   & 54 & L\_IPS1  & 99  & L\_10v    & 144 & L\_7m    \\
10 & L\_V3B   & 55 & L\_7AL   & 100 & L\_10pp   & 145 & L\_POS1  \\
11 & L\_LO1   & 56 & L\_7Am   & 101 & L\_11l    & 146 & L\_23d   \\
12 & L\_LO2   & 57 & L\_7PL   & 102 & L\_13l    & 147 & L\_v23ab \\
13 & L\_PIT   & 58 & L\_7PC   & 103 & L\_OFC    & 148 & L\_d23ab \\
14 & L\_MT    & 59 & L\_LIPv  & 104 & L\_TA2    & 149 & L\_31pv  \\
15 & L\_PreS  & 60 & L\_VIP   & 105 & L\_Pir    & 150 & L\_a24   \\
16 & L\_ProS  & 61 & L\_MIP   & 106 & L\_EC     & 151 & L\_d32   \\
17 & L\_PHA1  & 62 & L\_6v    & 107 & L\_H      & 152 & L\_p32   \\
18 & L\_PHA3  & 63 & L\_IFJp  & 108 & L\_PeEc   & 153 & L\_10r   \\
19 & L\_DVT   & 64 & L\_LIPd  & 109 & L\_STGa   & 154 & L\_47m   \\
20 & L\_V6A   & 65 & L\_6a    & 110 & L\_TGd    & 155 & L\_8Av   \\
21 & L\_VMV1  & 66 & L\_PFt   & 111 & L\_TE2a   & 156 & L\_8Ad   \\
22 & L\_VMV3  & 67 & L\_AIP   & 112 & L\_TF     & 157 & L\_9m    \\
23 & L\_PHA2  & 68 & L\_TE2p  & 113 & L\_25     & 158 & L\_8BL   \\
24 & L\_V4t   & 69 & L\_PHT   & 114 & L\_s32    & 159 & L\_9p    \\
25 & L\_V3CD  & 70 & L\_PH    & 115 & L\_pOFC   & 160 & L\_10d   \\
26 & L\_LO3   & 71 & L\_TPOJ3 & 116 & L\_TGv    & 161 & L\_45    \\
27 & L\_VMV2  & 72 & L\_PGp   & 117 & L\_PI     & 162 & L\_47l   \\
28 & L\_VVC   & 73 & L\_IP0   & 118 & L\_55b    & 163 & L\_a47r  \\
29 & L\_4     & 74 & L\_FST   & 119 & L\_RSC    & 164 & L\_9a    \\
30 & L\_3b    & 75 & L\_PSL   & 120 & L\_POS2   & 165 & L\_a10p  \\
31 & L\_A1    & 76 & L\_5mv   & 121 & L\_7Pm    & 166 & L\_47s   \\
32 & L\_5m    & 77 & L\_23c   & 122 & L\_33pr   & 167 & L\_s6-8  \\
33 & L\_5L    & 78 & L\_SCEF  & 123 & L\_8BM    & 168 & L\_AAIC  \\
34 & L\_24dd  & 79 & L\_6ma   & 124 & L\_8C     & 169 & L\_STSda \\
35 & L\_24dv  & 80 & L\_p24pr & 125 & L\_44     & 170 & L\_STSdp \\
36 & L\_1     & 81 & L\_a24pr & 126 & L\_IFJa   & 171 & L\_STSvp \\
37 & L\_2     & 82 & L\_p32pr & 127 & L\_IFSp   & 172 & L\_TE1a  \\
38 & L\_3a    & 83 & L\_6r    & 128 & L\_IFSa   & 173 & L\_PGi   \\
39 & L\_6d    & 84 & L\_43    & 129 & L\_p9-46v & 174 & L\_PGs   \\
40 & L\_6mp   & 85 & L\_52    & 130 & L\_46     & 175 & L\_31pd  \\
41 & L\_OP4   & 86 & L\_PFcm  & 131 & L\_a9-46v & 176 & L\_31a   \\
42 & L\_OP1   & 87 & L\_PoI2  & 132 & L\_9-46d  & 177 & L\_p10p  \\
43 & L\_OP2-3 & 88 & L\_FOP4  & 133 & L\_i6-8   & 178 & L\_STSva \\
44 & L\_RI    & 89 & L\_MI    & 134 & L\_AVI    & 179 & L\_TE1m  \\
45 & L\_PBelt & 90 & L\_FOP1  & 135 & L\_TE1p   & 180 & L\_p24   \\ \hline
\end{tabular}
\end{table}

\newpage
\bibliography{sn-bibliography.bib}